\documentclass[]{fairmeta_gr2}

\usepackage{amsmath, amssymb}
\usepackage{graphicx}
\usepackage{booktabs}
\usepackage{hyperref}
\usepackage{enumitem}
\usepackage[most]{tcolorbox}
\usepackage{subcaption}
\usepackage[dvipsnames]{xcolor}
\usepackage{colortbl}
\usepackage{pifont}
\usepackage{tcolorbox}
\usepackage{newunicodechar}
\newunicodechar{▷}{\texttriangleright}
\tcbuselibrary{listings, skins, breakable}
\usepackage{xcolor}
\usepackage{listings}
\usepackage[most]{tcolorbox}
\tcbuselibrary{listings,breakable}

\lstdefinelanguage{json}{
    morestring=[b]",
    morekeywords={true,false,null},
    keywordstyle=\color{orange!80!black}\bfseries,
    stringstyle=\color{green!45!black},
    showstringspaces=false,
    sensitive=true,
    literate=
        *{0}{{{\color{violet!80!black}0}}}{1}
         {1}{{{\color{violet!80!black}1}}}{1}
         {2}{{{\color{violet!80!black}2}}}{1}
         {3}{{{\color{violet!80!black}3}}}{1}
         {4}{{{\color{violet!80!black}4}}}{1}
         {5}{{{\color{violet!80!black}5}}}{1}
         {6}{{{\color{violet!80!black}6}}}{1}
         {7}{{{\color{violet!80!black}7}}}{1}
         {8}{{{\color{violet!80!black}8}}}{1}
         {9}{{{\color{violet!80!black}9}}}{1}
}

\newtcblisting{jsonblock}[3][]{
    enhanced,
    breakable,
    colback=#3!4,
    colframe=#3!60,
    colbacktitle=#3!75,
    coltitle=white,
    boxrule=0.5pt,
    arc=2pt,
    left=8pt,
    right=8pt,
    top=6pt,
    bottom=6pt,
    fonttitle=\bfseries\small,
    title={#2},
    listing only,
    listing options={
        language=json,
        basicstyle=\small\ttfamily,
        breaklines=true,
        breakatwhitespace=false,
        breakindent=0pt,
        breakautoindent=false,
        columns=fullflexible,
        keepspaces=true,
        showstringspaces=false,
        upquote=true,
        tabsize=2,
    },
    #1
}

\newtcolorbox{promptbox}[3][]{
    enhanced jigsaw,
    breakable,
    colback=#3!5,
    colframe=#3!60,
    boxrule=0.5pt,
    arc=2pt,
    left=8pt,
    right=8pt,
    top=6pt,
    bottom=6pt,
    fonttitle=\bfseries\small,
    fontupper=\small\ttfamily,
    title={#2},
    title after break={#2\space\textit{(continued)}},
    before upper={
        \setlength{\parindent}{0pt}
        \setlength{\parskip}{0pt}
        \obeylines
    },
    #1
}

\title{SCOReD: Student-Aware CoT Optimization for Recommendation Distillation}

\author[1]{Haz Sameen Shahgir}
\author[2,*]{Yufei Li}
\author[2]{Xiaohan Wei}
\author[2]{Yunchen Pu}
\author[2]{Fei Tian}
\author[2]{Chonglin Sun}
\author[2]{Frank Shyu}
\author[2]{Sandeep Pandey}
\author[2]{Luke Simon}
\author[1,\dagger]{Yue Dong}
\author[2,\dagger]{Xi Liu}

\affiliation[1]{University of California Riverside}
\affiliation[2]{Meta AI}

\contribution[*]{Execution lead}
\contribution[\dagger]{Joint correspon author}

\abstract{Chain-of-thought (CoT) distillation in the recommendation domain is a necessary precursor to RL training, but raw teacher traces are ill-suited to this task. Large teachers approach the recommendation task with unusually high reasoning uncertainty, repeatedly rechecking their answers without revising them; supervised fine-tuning on such traces produces verbose students that never revise their initial guess. Furthermore, due to the novelty of the recommendation domain, the teacher's reasoning traces are highly out-of-distribution for the small student LLM.

We propose \textbf{S}tudent-Aware \textbf{C}oT \textbf{O}ptimization for \textbf{Re}commendation \textbf{D}istillation (SCOReD), a CoT optimization framework tailored to recommendation that first parses each teacher trace into typed segments and uses the student LLM's attention to score the importance of each segment. Then SCOReD dynamically selects a per-segment edit (KEEP / REWRITE / FUSE / PRUNE) based on the output length and comparative log probability lift of the answer given the edit as per the student. Therefore, SCOReD prunes redundant sections of the reasoning trace while preserving information-dense sections and adapts raw teacher traces to the student's output distribution. Training on SCOReD-optimized CoTs provides a cleaner learning signal to the student model and improves over baseline SFT by $1.56\%$ NDCG and $1.9\%$ Recall@5, while reducing reasoning length by $27.3\%$.}

\date{\today}
\correspondence{Xi Liu (\email{xliu1@meta.com}) and Yue Dong (\email{yue.dong@ucr.edu})}

\begin{document}

\maketitle

\section{Introduction}
\label{section:intro}

Recent work has begun to recast recommendation as a generative reasoning problem rather than a purely discriminative ranking task. \citet{deng2025onerec} propose OneRec, an end-to-end generative recommender that unifies retrieval and ranking by directly generating recommendation outputs. Building on this direction, \citet{liu2025onerecthink} argue that generative recommenders should not operate only as implicit predictors, but should also expose explicit in-text reasoning over user intent and item semantics. \citet{liang2026gr2} similarly frame reranking as a generative reasoning task, where a model reasons over the user history and candidate items before producing the final ranked list. Together, these systems point toward a broader shift: recommendation models are increasingly expected not only to score items, but also to compare, justify, and revise candidate choices through language.

This shift creates a practical bottleneck. The strongest reasoning traces typically come from large LLMs, but real recommender systems often require small models because of latency, memory, and computational cost constraints. Distillation is therefore a necessary precursor for practical generative recommendation: before a small model can be improved with reinforcement learning or deployed as a reranker, it must first learn the basic format of recommendation reasoning from a stronger teacher. Prior work on chain-of-thought distillation shows that teacher rationales can transfer reasoning behavior to smaller models \citep{hsieh2023distilling,li2023symbolic}, while broader LLM distillation work shows that the distillation objective itself must be adapted for generative models \citep{gu2023minillm}.

\begin{figure}
    \centering
    \includegraphics[width=1\linewidth]{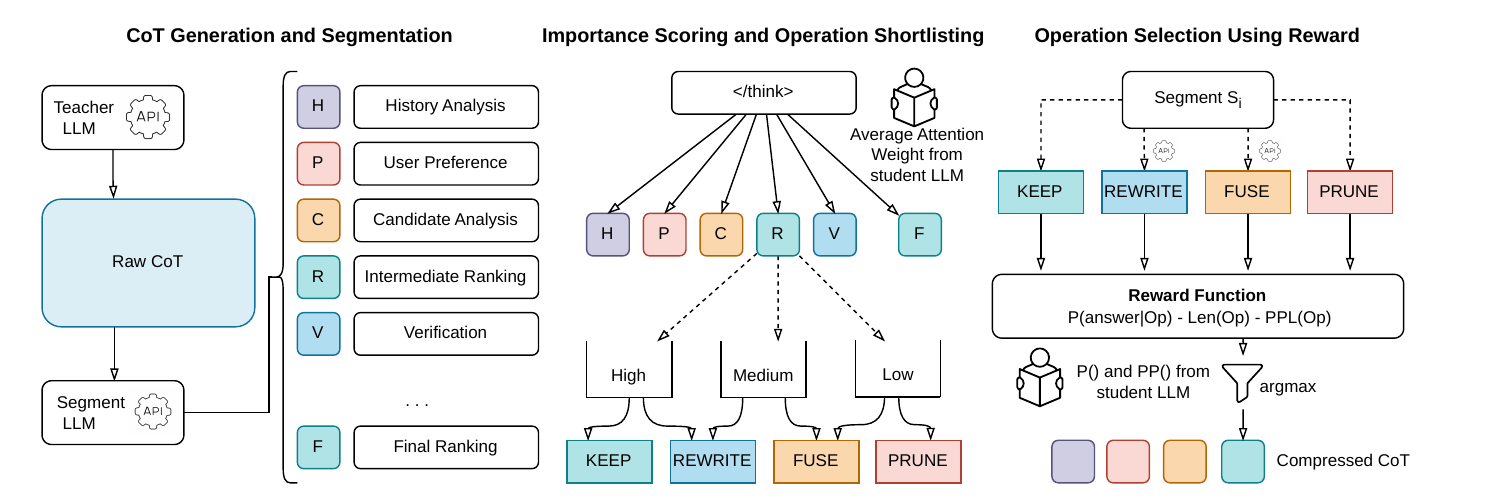}
    \caption{SCOReD recommendation trace compression pipeline. Left) Raw CoT from the teacher LLM is segmented into discrete spans and categorized into 6 stages. Middle) We compute the attention scores over the CoT using the pre-SFT target student LLM with a single forward pass. The average attention score from the \texttt{</think>} token approximates the student's perceived importance of each stage and is used to shortlist modification operations. Right) The final operation is selected based on a reward function that accounts for the student's likelihood, perplexity, and the length of the resulting text. }
    \label{fig:open_diagram}
\end{figure}

Recommendation, however, makes CoT distillation unusually difficult. Unlike math or coding, recommendation labels come from noisy user behavior rather than expert-verified derivations. A user may purchase an item that only weakly follows from their history, ignore an apparently relevant item, or express multiple competing interests in the same session. As a result, recommendation does not usually have a single clean reasoning path. The teacher's rationale is often a negotiation among ambiguous preferences, overlapping candidates, and noisy labels. This produces a dominant pathology: \textbf{repetitive verification that rarely changes the final answer}. The teacher repeatedly revisits the same candidate decisions after already forming an answer. On average, it rechecks its initial answer $3.2$ times per trace, and in $79\%$ of cases these rechecks do not change the final answer. They simply restate the same rationale in slightly different words. Under naive supervised fine-tuning, this becomes the dominant signal given to the student: reason verbosely, repeatedly verify, and rarely revise.

Recent work on CoT compression shows that long rationales often contain redundant tokens and can be shortened for efficiency. C3oT trains models to generate shorter chains of thought while preserving reasoning performance \citep{kang2024c3ot}; TokenSkip learns to skip less important CoT tokens under controllable compression ratios \citep{xia2025tokenskip}; CRISP uses intrinsic attention saliency around the reasoning termination token to prune redundant reasoning \citep{lan2026crisp}; and Compress-Distill uses an LLM-based compressor before distillation \citep{griot2026compressdistill}. However, these works focus on domains such as mathematics, coding, and science, where the ground-truth answer is fixed, and CoT does not exhibit the redundancy we observe in the recommendation domain. Compression CoT risks excising crucial logical steps, rendering the resulting compressed CoT out-of-distribution for the student model. As a result, all prior compression methods have traded off performance for shorter CoTs. Across all prior work, training the student on the raw uncompressed CoT proved to be the most performant.

We build on CRISP's \citep{lan2026crisp} attention-based CoT compression to accurately excise redundant reasoning steps of the teacher LLM while simultaneously ensuring the compressed CoT is in-distribution for the student LLM. SCOReD first segments each teacher trace into recommendation-specific stages such as purchase-history review, preference modeling, candidate analysis, verification, and final ranking. It then scores each segment using the attention from the \texttt{</think>} token to the segment's tokens using the target student LLM. Based on this score, SCOReD shortlists from the following operations: \textsc{Keep}, \textsc{Rewrite}, \textsc{Fuse}, or \textsc{Prune}. The final operation is determined by using a reward function based on the log-probability of the student generating the answer conditioned on the operation and on the perplexity. This scheme allows SCOReD to improve over the original trace (such as in cases where P(answer|Rewrite(segment)) $>$ P(answer|Keep(segment)) while also ensuring the operation is in-distribution for the student LLM.

The result is a cleaner supervision signal for small recommendation models. Instead of imitating the teacher's repetitive uncertainty, the student learns shorter rationales that preserve decision-relevant candidate comparisons while removing redundant rechecks. Empirically, we show for the first time that SCOReD-optimized CoT can surpass the use of raw traces in the recommendation domain. Training on SCOReD-optimized CoT improves over raw-trace SFT by 1.56\% NDCG and 1.9\%
Recall@5, while reducing reasoning length by 27.3\%.

\section{Related Work}
\label{sec}

\paragraph{Generative recommendation.}
Traditional recommendation pipelines retrieve candidates and then score them with separate models, making it difficult to generate explanations, rankings, or recommendations through a unified reasoning process. Generative recommendation addresses this limitation by casting recommendation as sequence generation, allowing a single model to directly produce recommendation outputs. P5 \citep{geng2022p5} unifies rating prediction, sequential recommendation, and explanation as personalized text-to-text tasks, while M6-Rec \citep{cui2022m6rec} adapts an industrial pretrained model across retrieval, ranking, explanation, and content generation with efficient tuning. TIGER \citep{rajput2023tiger} instead learns semantic item identifiers for autoregressive retrieval, whereas LLMRank \citep{hou2024llmrank} prompts LLMs to perform zero-shot listwise reranking. OneRec \citep{deng2025onerec} further replaces the retrieve--rank cascade with a unified generator and iterative preference alignment. Reasoning-aware systems extend this line: OneRec-Think \citep{liu2025onerecthink} grounds item tokens in text, activates explicit reasoning through scaffolding, and uses rewards designed for multiple valid recommendations; GR2 \citep{liang2026gr2} mid-trains on semantic IDs, distills teacher-generated reranking rationales through SFT, and applies DAPO with verifiable ranking rewards. Both rely on obtaining reasoning traces that match logged recommendation targets despite the fact that recommendation labels are noisy and often non-unique. A common solution is targeted generation, where the teacher is conditioned to explain a predetermined item. While this increases label alignment, it also encourages post-hoc rationalization: the teacher is no longer explaining its own decision process, but constructing a justification for an externally specified outcome.

\paragraph{CoT distillation.}
CoT prompting \citep{wei2022chain} showed that intermediate steps can improve large-model reasoning, while self-consistency \citep{wang2023selfconsistency} aggregates diverse sampled paths instead of trusting one greedy derivation. Distillation turns this behavior into supervision: Distilling Step-by-Step \citep{hsieh2023distilling} jointly trains on labels and teacher rationales; Symbolic CoT Distillation \citep{li2023symbolic} and Teaching Small Language Models to Reason \citep{magister2023teaching} show that much smaller students can acquire stepwise reasoning from sampled traces; and MiniLLM \citep{gu2023minillm} uses reverse-KL sequence-level distillation to reduce mode averaging and exposure bias. Distillation also plays a central role in recent reasoning-based recommendation systems: both OneRec-Think \citep{liu2025onerecthink} and GR2 \citep{liang2026gr2} first transfer reasoning behavior from a large teacher into a smaller recommender through supervised distillation before applying reinforcement learning. However, recommendation differs from traditional reasoning tasks because logged labels are noisy and represent only one of many plausible items. As a result, teachers may not naturally predict the target item, motivating targeted generation or filtering to obtain label-compatible traces. These procedures can introduce rationalization, selection bias, or traces that are difficult for the student to learn, making recommendation-specific CoT distillation an important but still underexplored problem.

\paragraph{CoT compression.}
Reasoning compression operates at prompt, token, step, and trajectory levels. LLMLingua \citep{jiang2023llmlingua} performs budgeted coarse-to-fine token pruning; C3oT \citep{kang2024c3ot} jointly trains on long and compressed traces; and TokenSkip \citep{xia2025tokenskip} learns controllable shortcuts by dropping low-importance tokens. Step Entropy \citep{li2025stepentropy} removes low-information steps, whereas CRISP \citep{lan2026crisp} uses model-internal saliency to prune redundancy. For distillation, Compress-Distill \citep{griot2026compressdistill} compresses correct teacher traces post hoc before SFT, and mixed-policy distillation \citep{yang2026mpd} has a teacher rewrite student-sampled trajectories, preserving the student's policy while transferring concise behavior. Existing compression methods generally view reasoning traces as correct and seek shorter versions that preserve as much performance as possible. In practice, however, compression almost always incurs some accuracy degradation because shortening a trace can remove steps that are important for a downstream model's reasoning process, alter the logical structure of the derivation, or produce compressed traces that are difficult for the target student to model. Since compression is typically performed without considering the student's capabilities, the resulting traces may also contain high-perplexity tokens or transitions that are out-of-distribution for the learner. Recent work shows that learning from high-perplexity tokens can contribute to catastrophic forgetting during fine-tuning \citep{wu2025mitigatingforgettingllmfinetuning}.

\section{SCOReD Method}
\label{sec:method}

SCOReD consists of three stages: LLM-based chain-of-thought (CoT) segmentation, attention-based importance scoring, and per-segment edit selection based on each segment's importance score and relative advantage under a reward function.

\subsection{LLM-Based CoT Segmentation}
\label{sec:segmentation}
We first generate $K_{rollout}$ reasoning traces using the teacher LLM, Google Gemma-4-26B, with the recommendation prompt shown below.

\begin{promptbox}{Candidate Ranking System Prompt}{blue}
You are a recommendation assistant. You will receive a user's purchase history and a numbered list of candidate items. Your task is to rank all candidates from most to least likely to match the user's preferences, where preference is inferred from thematic, categorical, and ingredient/style similarity to past purchases.
\medskip
\textbf{Output format:} Return only a Python-style list containing every candidate number exactly once, ordered from most to least preferred. Example for 10 candidates: [3, 6, 7, 2, 4, 9, 10, 5, 1, 8]. Do not include explanations, commentary, or any text outside the list in the final answer.
\end{promptbox}

Our procedure is similar to the rejection-sampling approach of \citet{liang2026gr2}. However, rather than requiring an exact match with the ground-truth ranking, we select the CoT associated with the highest NDCG score. Ties are broken in favor of the shorter CoT.

The recommendation prompt constrains only the format of the final output and imposes no structural requirements on the teacher model's reasoning process. This is important because \citet{yuehhan2026reasoningmodelsstrugglecontrol} show that reasoning LLMs successfully control the structure of their CoT in only $2.7\%$ of cases. We therefore formulate structured reasoning as a post-hoc extraction problem and use a second LLM pass to identify the underlying reasoning stages.

After generating the traces, we use Gemma-4-26B to partition each CoT into a sequence of contiguous segments, $S_1, \dots, S_n$, using the prompt specified in Appendix \ref{app:cot_segment_prompt}. Each segment is assigned one of six labels: Purchase History Review, User Interest Modeling, Candidate Analysis, Intermediate Ranking, Verification, or Final Ranking.

Table~\ref{tab:stage-stats} reports segment-level statistics for $289{,}541$ annotated blocks from $11,350$ CoT traces in our training set. Verification (\textbf{V}) and ranking (\textbf{R}) are the most common stages and together account for approximately $63\%$ of all segments. Final-ranking segments (\textbf{F}) are considerably shorter on average, consistent with their role as concise concluding statements.

\begin{table}[t]
\centering
\caption{Segmentation stage distribution (segment-level counts).}
\label{tab:stage-stats}
\begin{tabular}{lrrr}
\toprule
Stage & Share & Avg.\ chars/seg & Role \\
\midrule
V (Verification)      & 33.9\% & 418 & Pairwise checks, doubts, revisions \\
R (Ranking)           & 29.4\% & 367 & Tiering, draft orderings \\
C (Candidate Analysis)& 15.3\% & 459 & Per-candidate relevance \\
F (Final Ranking)     &  7.3\% & 120 & Final ordered conclusions \\
P (Preference)        &  7.1\% & 169 & Theme / preference inference \\
H (Purchase History)  &  7.0\% & 268 & History summarization \\
\bottomrule
\end{tabular}
\end{table}

\begin{table}[t]
\centering
\caption{Adjacent stage transition counts $N_{ij}$ (blue heatmap; max cell $=70{,}414$).}
\label{tab:transition-counts}
\small
\setlength{\tabcolsep}{4pt}
\begin{tabular}{l|rrrrrr}
\toprule
From $\backslash$ To & H & P & C & R & V & F \\
\midrule
H & \cellcolor{white}{0} & \cellcolor{blue!17}{12{,}130} & \cellcolor{blue!8}{4{,}204} & \cellcolor{blue!8}{1{,}343} & \cellcolor{blue!8}{2{,}620} & \cellcolor{blue!8}{104} \\
P & \cellcolor{blue!8}{126} & \cellcolor{white}{0} & \cellcolor{blue!20}{13{,}943} & \cellcolor{blue!8}{4{,}343} & \cellcolor{blue!8}{1{,}736} & \cellcolor{blue!8}{321} \\
C & \cellcolor{blue!8}{912} & \cellcolor{blue!8}{2{,}475} & \cellcolor{white}{0} & \cellcolor{blue!34}{23{,}749} & \cellcolor{blue!22}{15{,}436} & \cellcolor{blue!8}{1{,}558} \\
R & \cellcolor{blue!8}{2{,}390} & \cellcolor{blue!8}{1{,}345} & \cellcolor{blue!11}{7{,}862} & \cellcolor{white}{0} & \cellcolor{blue!100}{\textcolor{white}{70{,}414}} & \cellcolor{blue!8}{2{,}337} \\
V & \cellcolor{blue!8}{4{,}768} & \cellcolor{blue!8}{4{,}415} & \cellcolor{blue!25}{17{,}433} & \cellcolor{blue!74}{\textcolor{white}{52{,}407}} & \cellcolor{white}{0} & \cellcolor{blue!24}{16{,}687} \\
F & \cellcolor{blue!8}{880} & \cellcolor{blue!8}{95} & \cellcolor{blue!8}{849} & \cellcolor{blue!8}{3{,}233} & \cellcolor{blue!11}{8{,}076} & \cellcolor{white}{0} \\
\bottomrule
\end{tabular}
\end{table}

The segmented traces reveal several dominant structural patterns. The teacher's CoT consistently begins with the sequence $\textbf{H} \rightarrow \textbf{P} \rightarrow \textbf{C} \rightarrow \textbf{R}$. In the \textbf{H} stage, the teacher reviews and categorizes the user's purchase history, identifying recurring patterns among previously purchased items. It then uses these observations in the \textbf{P} stage to infer the user's interests and preferences. Next, the teacher evaluates the candidate items in the \textbf{C} stage by comparing them with the purchase history and inferred preferences, before producing an initial ranking in the \textbf{R} stage.

Following this initial ranking, the teacher often performs several refinement loops. These loops are commonly characterized by transitions of the form $\textbf{R} \rightarrow \textbf{V} \rightarrow \textbf{R}$ or $\textbf{R} \rightarrow \textbf{V} \rightarrow \textbf{C} \rightarrow \textbf{R}$. On average, a raw teacher CoT contains $8.65 \pm 4.17$ verification stages. This frequency highlights the substantial uncertainty exhibited even by a capable teacher model in the recommendation domain.

Despite containing an average of $8.65$ verification stages, only $47.15\%$ of traces produce a final ranking that differs from the initial ranking. Moreover, $92.89\%$ of individual refinement loops leave the proposed ranking unchanged. Consequently, training a student directly on the raw traces may not teach meaningful backtracking or refinement. Instead, it may primarily teach the student to repeatedly reinforce its initial ranking while substantially increasing the length of the CoT.

\paragraph{Is the Number of Refinement Attempts Predictable?}
We next investigate whether simple lexical properties of the input can predict how many refinement attempts the teacher will perform. For example, a longer purchase history might plausibly induce additional verification steps. However, we find that prompt length, purchase-history length, and textual overlap between purchased and candidate items are only weakly correlated with the number of \textbf{V} stages, with Pearson correlation coefficients of $-0.08$, $-0.05$, and $-0.11$, respectively. These results suggest that the teacher's refinement behavior is highly context-dependent and cannot be reliably predicted from simple lexical features, further illustrating the difficulty of recommendation reasoning for LLMs.

\subsection{Importance Scoring via \texttt{</think>}-Attention}
\label{sec:scoring}

We assign each segment an importance score based on the average attention its tokens receive from the \texttt{</think>} delimiter that terminates the CoT. \citet{lan2026crisp} observe that tokens generated \emph{after} \texttt{</think>} attend almost exclusively to the delimiter itself rather than directly to the preceding reasoning trace. This behavior suggests that \texttt{</think>} functions as a learned summary representation of the CoT. We therefore use the average attention from \texttt{</think>} to the tokens in a segment as a proxy for that segment's contribution to the final answer.

Based on these scores, we partition segments into three importance buckets: \textsc{Low}, containing the bottom $f_{low}\%$ of segments by attention score; \textsc{High}, containing the top $f_{high}\%$; and \textsc{Medium}, containing the remaining segments.

\subsection{Reward-based Action Selection}
Each importance bucket defines a restricted action set: \textsc{High} segments may be \textsc{Kept} or \textsc{Rewritten}, \textsc{Medium} segments may be \textsc{Rewritten} or \textsc{Fused}, and \textsc{Low} segments may be \textsc{Fused} or \textsc{Pruned}. Let $\mathcal{A}_i$ denote the set of actions available for segment $S_i$. The \textsc{Rewrite} and \textsc{Fuse} actions are implemented using an LLM, whereas \textsc{Keep} and \textsc{Prune} are applied directly.

For each action $a \in \mathcal{A}_i$, let $\widetilde{S}_i^a = a(S_i)$ denote the resulting segment, and let $C_i = (prompt, S_{1:i-1})$ denote the prompt together with the previously processed segments. We evaluate each candidate edit using

\begin{equation}
\label{eq:reward_2}
R_i(a)
=
\log P(answer \mid C_i, \widetilde{S}_i^a)
-
\alpha\,Len(\widetilde{S}_i^a)
-
\beta\,PPL(\widetilde{S}_i^a \mid C_i).
\end{equation}

The term $\log P(answer \mid C_i, \widetilde{S}_i^a)$ is the student model's log probability of the teacher's final ranking answer after conditioning on the prompt, the preceding segments, and the candidate edit. $Len(\widetilde{S}_i^a)$ is the number of tokens in the edited segment and penalizes unnecessarily long reasoning. $PPL(\widetilde{S}_i^a \mid C_i)$ is the perplexity of the edited segment under the student model, which penalizes edits that are difficult for the student to model. The coefficients $\alpha$ and $\beta$ control the strengths of the length and perplexity penalties, respectively. We then select the highest-reward action for each segment:

\begin{equation}
a_i^* = \arg\max_{a \in \mathcal{A}_i} R_i(a).
\end{equation}

In our implementation, we set the importance thresholds $f_{low}=f_{high}=10\%$. We use Gemma-4-26B for \textsc{Rewrite} and \textsc{Fuse}. In our reward function, we set length penalty weight $\alpha=0.005$ and perplexity penalty weight $\beta=0.1$.

\subsection{Baseline: LLM-Based Summarization}
Recent work has explored post-hoc reasoning-trace compression as a means of reducing the training and inference computational cost of knowledge distillation, while observing that compression introduces an accuracy--efficiency trade-off rather than a uniform improvement \citep{griot2026compressdistill}. Following this approach, we include a simple one-shot summarization baseline. Given the segmented trace from Section~\ref{sec:segmentation}, an LLM receives the complete trace and rewrites it in a single pass. The prompt instructs the model to consolidate repeated candidate analysis, remove redundant verification, and retain revisions that reflect meaningful backtracking.

\begin{promptbox}{LLM Summarizer System Prompt}{blue}
You are a distillation agent. Given the teacher's original reasoning chain, you will rewrite it to be more pedagogical for a student model. You will:
\medskip
1) Consolidate reasoning about candidates.
2) Remove redundant doubts that repeat a previous step's reasoning.
3) Keep refinement/doubts that are logical. If the intermediate verdict differs from the final answer, keep both to teach the student model backtracking.
4) Do not shorten Purchase History Analysis, User Preference Modeling, initial Candidate Analysis, or Final Ranking.
5) You may enrich Purchase History Analysis, User Preference Modeling, and initial Candidate Analysis by adding more details from later steps.
6) Do not change the name of the steps.
\end{promptbox}

Unlike SCOReD, which processes segments from left to right and fuses only adjacent segments, the one-shot baseline receives the entire trace at once. It is therefore not subject to the same locality constraint, although its decisions about which segments to remove, retain, or combine depend entirely on the summarizer's judgment. The baseline also requires only one LLM generation per trace, making it less computationally expensive than evaluating multiple candidate edits for each segment.

However, the summarizer is not optimized for the student model. It does not use attention-based segment importance, the student's probability of producing the target answer, or the perplexity of the edited reasoning. Consequently, it may remove reasoning that is useful to the student or preserve content that is fluent but redundant. In contrast, SCOReD selects edits using segment-level signals derived directly from the student model.

\subsection{Post-SFT Optimization using Reinforcement Learning and On-Policy Distillation}
\label{sec:post-sft-optimization}

Supervised fine-tuning transfers the structure of the teacher's reasoning into the student, but it optimizes token likelihood rather than recommendation quality under the student's own generation distribution. We therefore investigate two complementary forms of post-SFT optimization. First, we apply reinforcement learning with a sequence-level ranking reward using DAPO \citep{yu2026dapo}. Second, we apply On-Policy Self-Distillation (OPSD) \citep{opsd}, which provides dense token-level supervision on trajectories sampled from the student. Both methods are initialized from the student trained on SCOReD-optimized CoTs.

Prior reasoning-based recommenders have primarily used reinforcement learning to optimize task performance. OneRec-Think applies GRPO with a recommendation-specific rollout-beam reward designed to reduce reward sparsity in generative retrieval \citep{liu2025onerecthink}, while GR2 applies DAPO with verifiable reranking rewards \citep{liang2026gr2}. OneRec also incorporates on-policy distillation, but uses it on general-domain prompts to restore instruction-following and reasoning capabilities degraded during recommendation training; recommendation performance is subsequently optimized through a separate reinforcement-learning stage. In contrast, we evaluate whether both reinforcement learning and on-policy distillation can directly improve recommendation performance after CoT distillation.

\paragraph{Reinforcement Learning with DAPO.}
Let $\pi_{\theta_0}$ denote the SCOReD SFT policy. For each recommendation prompt $x$, we sample a group of responses
$$
y_1,\ldots,y_G \sim \pi_{\theta}(\cdot \mid x).
$$
We parse the final ranking in each response and compute a verifiable sequence-level reward:
\begin{equation}
r(y_g,y^*)
=
\begin{cases}
\operatorname{NDCG}(y_g,y^*), & \text{if $y_g$ is a valid ranking},\\
0, & \text{otherwise},
\end{cases}
\label{eq:dapo-reward}
\end{equation}
where $y^*$ denotes the set of logged relevant items. DAPO constructs relative advantages from the rewards of responses sampled for the same prompt and increases the likelihood of trajectories that obtain higher ranking quality. This objective directly aligns the policy with the test-time ranking metric while penalizing malformed outputs through their zero reward. We initialize both the policy and reference model from $\pi_{\theta_0}$ and perform LoRA-based optimization to limit deviation from the SCOReD SFT policy.

\paragraph{On-Policy Self-Distillation with Privileged Information.}
Standard On-Policy Distillation (OPD) requires a teacher that is stronger than the student while sharing its vocabulary, because the teacher and student token distributions are compared along student-generated trajectories \citep{opd}. Gemma-4-26B provides a stronger recommendation policy in our setting, but its vocabulary is incompatible with that of the Qwen student. Conversely, Qwen-3.6-35B-A3B is vocabulary-compatible but does not provide a consistently stronger policy on our reranking task. We therefore use On-Policy Self-Distillation with privileged information (OPSD), in which the same Qwen model acts as both teacher and student under different conditioning contexts.

For each training instance, the logged ground truth identifies a set of relevant candidates rather than a complete ordering of all candidates. We therefore construct a privileged reference ranking $\widetilde{y}^{*}$ by stably moving the ground-truth items to the beginning of the teacher ranking while preserving the relative order of the remaining candidates. The student policy observes only the original recommendation prompt,
$$
\pi_{\theta}^{S}(\cdot \mid x),
$$
whereas the privileged teacher policy additionally observes $\widetilde{y}^{*}$,
$$
\pi^{T}(\cdot \mid x,\widetilde{y}^{*}).
$$

The student first samples an on-policy response $y \sim \pi_{\theta}^{S}(\cdot \mid x)$. At each position $t$, both policies then evaluate the same student-generated prefix $y_{<t}$. We minimize the divergence between their token distributions:
\begin{equation}
\mathcal{L}_{\mathrm{OPSD}}
=
\mathbb{E}_{x}
\mathbb{E}_{y\sim\pi_{\theta}^{S}(\cdot\mid x)}
\left[
\frac{1}{|y|}
\sum_{t=1}^{|y|}
\mathcal{D}
\left(
\operatorname{sg}\!\left[
\pi^{T}(\cdot\mid x,\widetilde{y}^{*},y_{<t})
\right],
\pi_{\theta}^{S}(\cdot\mid x,y_{<t})
\right)
\right],
\label{eq:opsd-objective}
\end{equation}
where $\mathcal{D}$ denotes the token-distribution divergence and $\operatorname{sg}$ denotes stop-gradient through the privileged teacher distribution. OPSD therefore trains on prefixes encountered under the student's own policy while using the ground-truth items to provide dense guidance about the desired continuation.

DAPO and OPSD provide complementary post-SFT signals. DAPO directly optimizes the non-differentiable ranking metric, but assigns a single sequence-level reward to the complete trajectory. OPSD instead supplies dense token-level supervision, although its effectiveness depends on whether the small student can exploit the privileged ranking context. We evaluate whether either approach improves upon the SCOReD SFT initialization in Section~\ref{sec:post-sft-results}.

\section{Experimental Setup}
\label{sec:experimental_setup}

\paragraph{Dataset Construction}
We create our recommendation dataset from Amazon Beauty Pretrain\footnote{\url{https://github.com/wangshy31/OneRec-Think/tree/main/data}}~\citep{hou2026bridginglanguageitemsretrieval}, which contains raw user interaction logs. We filter out all samples with fewer than 5 items. Given a log of N items, we take the last 3 items as the ground truth to be predicted from the previous N-3 historic items. The length of the resultant purchase history ranges from 2 to 46, with a median of 6. We randomly sample $7$ negative candidates from the Amazon Beauty Pretrain item database, resulting in $K_{candidate}=10$ for all samples. We use a 60:20:20 data split, which yields 13417 training samples, 4472 validation samples, and 4474 test samples. After LLM-based CoT segmentation, we filtered out unparseable samples, resulting in a final training set of 11,350. 

\paragraph{Model Selection}
We use Google Gemma-4-26B-A4B-it \footnote{\url{google/gemma-4-26B-A4B-it}} as the teacher LLM and Qwen-3-0.6 \footnote{\url{https://huggingface.co/Qwen/Qwen3-0.6B}} as the student. We use Gemma-4-26B for CoT segmentation and also the REWRITE and FUSE operations in our compression algorithm.

\paragraph{Model Training}
For Supervised Fine-Tuning, we use full-parameter finetuning using the \texttt{verl} library with a batch size of 64 and 10 epochs with early stopping. We sweep the learning rate from $1e-6$ to $1e-3$, and select the best one using the validation set. For all training regimes, LR=$3e-5$ proved to be optimal. For DAPO \citep{yu2026dapo} and OPSD \citep{opsd}, we use LoRA ($r=32$, $\alpha=32$) finetuning due to compute constraints. For DAPO, we use a batch of 16 prompts, with 8 rollouts per prompt and train for 6 epochs with learning rate $1e-5$. For OPSD, we use a batch size of 32 and $LR=5e-6$ and train for 3200 samples before early stopping.

\section{Results}
\label{sec:results}

\label{sec:main-results}

\begin{table}[!h]
\centering
\caption{Reranking results on all $4{,}474$ test samples. Parse failures receive a score of zero. Gemma-4-26B is the teacher used to generate the training traces. \textbf{Bold} and \underline{underlining} denote the best and second-best results among trained students, respectively.}
\label{tab:rerank-large-test-all}
\begin{tabular}{lrrr|rr}
\toprule
 & \multicolumn{3}{c}{\textbf{Trained SLMs (Qwen3-0.6B)}}
 & \multicolumn{2}{c}{\textbf{Reference LLMs}} \\
\cmidrule(lr){2-4}
\cmidrule(lr){5-6}
 & \textbf{Baseline}
 & \textbf{LLM-Summ.}
 & \textbf{SCOReD}
 & \textbf{Gemma-4}
 & \textbf{Qwen-3.6} \\
\midrule
Model size
    & 0.6B & 0.6B & 0.6B & 26B-E4B & 35B-A3B \\
Training
    & SFT & SFT & SFT & --- & --- \\
\midrule
Parse failures
    & \underline{126} & 391 & \textbf{68} & 0 & 94 \\
Parse failure rate (\%)
    & \underline{2.82} & 8.74 & \textbf{1.52} & 0.00 & 2.10 \\
Test samples
    & 4474 & 4474 & 4474 & 4474 & 4474 \\
Trace length (K chars)
    & 8.5$\pm$2.9
    & 4.4$\pm$3.9
    & 6.2$\pm$1.7
    & 10.0$\pm$3.7
    & 9.4$\pm$2.1 \\
\midrule
NDCG
    & \underline{0.7786}
    & 0.7342
    & \textbf{0.7908}
    & 0.8030
    & 0.7879 \\
MRR
    & \underline{0.7592}
    & 0.7179
    & \textbf{0.7725}
    & 0.7827
    & 0.7683 \\
MAP
    & \underline{0.6558}
    & 0.6197
    & \textbf{0.6667}
    & 0.6781
    & 0.6666 \\
Hit@1
    & \underline{0.6341}
    & 0.6021
    & \textbf{0.6453}
    & 0.6560
    & 0.6455 \\
Hit@3
    & \underline{0.8657}
    & 0.8203
    & \textbf{0.8865}
    & 0.8929
    & 0.8735 \\
Hit@5
    & \underline{0.9435}
    & 0.8920
    & \textbf{0.9564}
    & 0.9689
    & 0.9464 \\
Recall@1
    & \underline{0.2114}
    & 0.2007
    & \textbf{0.2151}
    & 0.2187
    & 0.2152 \\
Recall@3
    & \underline{0.5335}
    & 0.5063
    & \textbf{0.5445}
    & 0.5527
    & 0.5443 \\
Recall@5
    & \underline{0.7108}
    & 0.6746
    & \textbf{0.7243}
    & 0.7366
    & 0.7221 \\
Precision@1
    & \underline{0.6341}
    & 0.6021
    & \textbf{0.6453}
    & 0.6560
    & 0.6455 \\
Precision@3
    & \underline{0.5335}
    & 0.5063
    & \textbf{0.5445}
    & 0.5527
    & 0.5443 \\
Precision@5
    & \underline{0.4265}
    & 0.4047
    & \textbf{0.4346}
    & 0.4419
    & 0.4333 \\
\bottomrule
\end{tabular}%

\end{table}

Table~\ref{tab:rerank-large-test-all} reports reranking performance on all $4{,}474$ test samples. Outputs that cannot be parsed as valid rankings receive a score of zero, so the reported metrics jointly measure ranking quality and output-format reliability.

SCOReD achieves the best performance among the trained $0.6$B students on every reported ranking metric. Relative to standard SFT on the uncompressed teacher traces, SCOReD improves NDCG from $0.7786$ to $0.7908$, MRR from $0.7592$ to $0.7725$, and MAP from $0.6558$ to $0.6667$. These improvements are obtained while reducing the average trace length from $8.5$K to $6.2$K characters, corresponding to a reduction of approximately $27\%$. SCOReD also reduces the parse-failure rate from $2.82\%$ to $1.52\%$, a relative reduction of approximately $46\%$. Despite using only $0.6$B parameters, our student outperforms Qwen-3.6-35B-A3B and approaches the performance of the Gemma-4-26B teacher. Table \ref{tab:parse-failure-taxonomy} shows that the majority of the errors arise from the models generating a final ranking with fewer than the required $K$ candidates specified in the user's prompt. LLM-based compression also induces more Duplicate Indices failures where the final ranking contains the same candidate more than once. This shows that although LLM-based compression achieves the highest CoT length reduction, this compressed CoT is highly out-of-distribution for the student model and interferes with simple capabilities such as producing a list with unique elements.

\begin{table}[t]
\centering
\caption{Parse-failure taxonomy on the $4{,}474$-sample reranking test with $k=1$. Categories are mutually exclusive, and percentages are calculated within each model's parse failures.}
\label{tab:parse-failure-taxonomy}
\small
\begin{tabular}{lrrr}
\toprule
\textbf{Failure type}
    & \textbf{Baseline SFT} & \textbf{LLM-Compression} & \textbf{SCOReD} \\
    & (126 failures)& (391 failures) & (68 failures) \\
\midrule
Incomplete list ($<K$ indices)& 112 (88.9\%)& 289 (73.9\%)& 54 (79.4\%) \\
Duplicate indices& 1 (0.8\%)& 72 (18.4\%)& 7 (10.3\%) \\
No final list& 6 (4.8\%)& 8 (2.0\%)& 0 (0.0\%) \\
Product names / invalid JSON\textsuperscript{*}& 0 (0.0\%)& 15 (3.8\%)& 1 (1.5\%) \\
Invalid index values& 4 (3.2\%)& 5 (1.3\%)& 2 (2.9\%) \\
List too long ($>10$ indices)& 3 (2.4\%)& 2 (0.5\%)& 4 (5.9\%) \\
\midrule
Total parse failures& 126 (2.8\%)& 391 (8.7\%)& 68 (1.5\%) \\
\bottomrule
\end{tabular}

\vspace{0.4em}
{\footnotesize
\textsuperscript{*}The final answer contains bracketed content that is not a valid list of integer indices, such as product names \texttt{[Cover Girl Concealer, Cream]} or mixed tokens \texttt{[3, C2, C5, \ldots]}.}
\end{table}

\begin{figure}[t]
    \centering
    \begin{subfigure}[b]{0.48\textwidth}
        \centering
        \includegraphics[width=\textwidth]
        {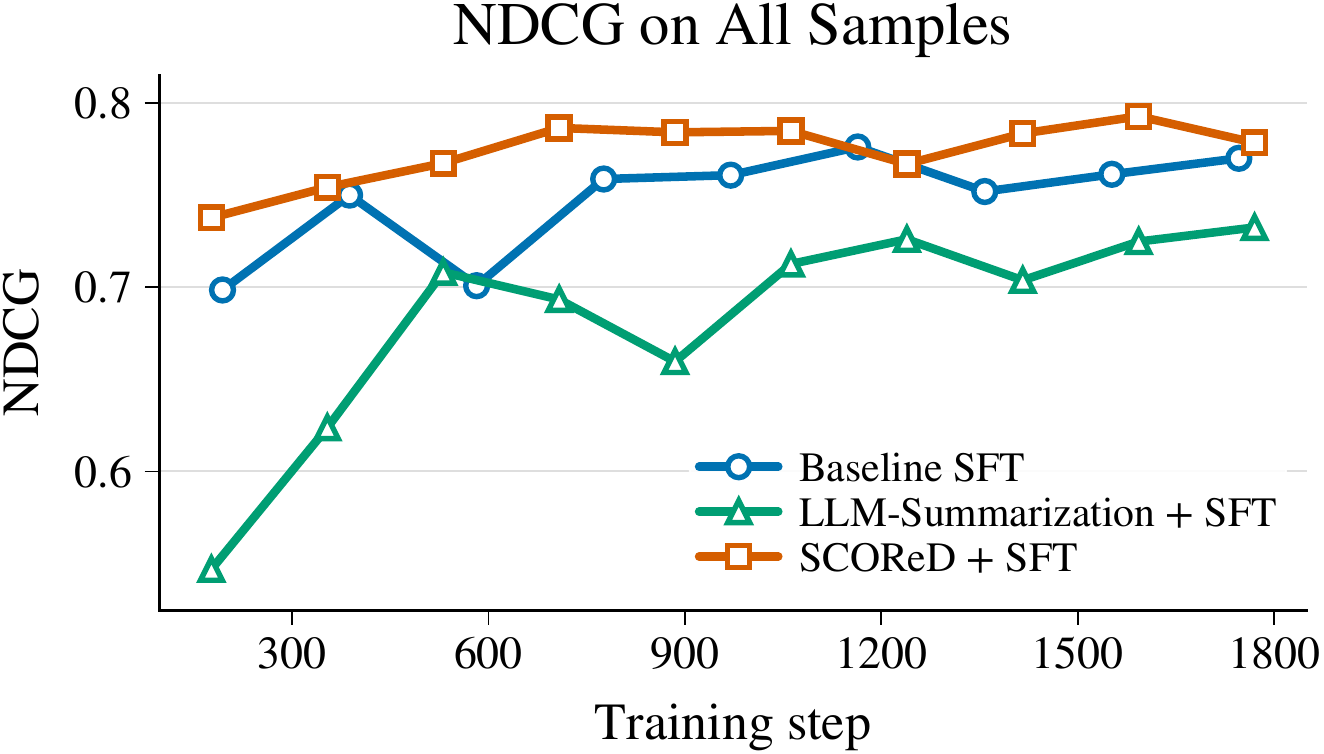}
        \caption{NDCG on all test samples.}
        \label{fig:ndcg-all-training}
    \end{subfigure}
    \hfill
    \begin{subfigure}[b]{0.48\textwidth}
        \centering
        \includegraphics[width=\textwidth]
        {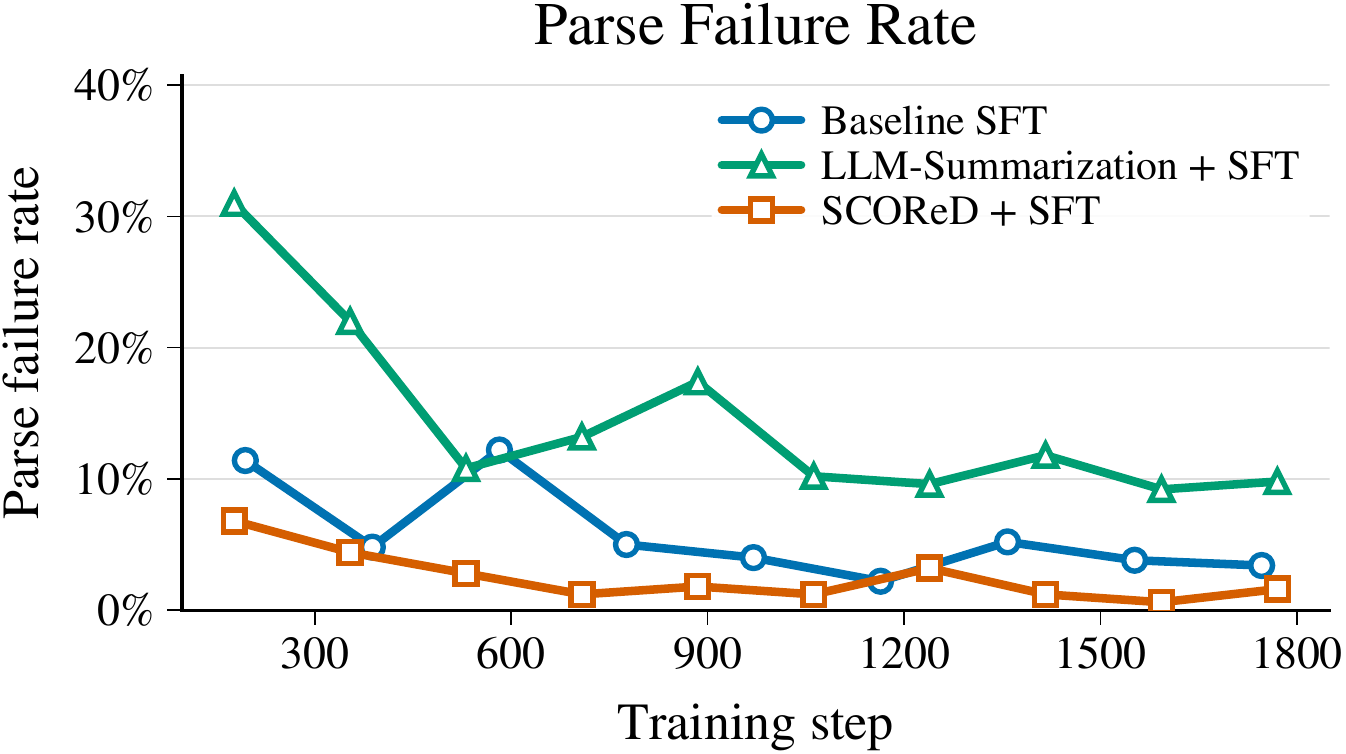}
        \caption{Parse-failure rate.}
        \label{fig:parse-failure-training}
    \end{subfigure}
    \caption{All-sample ranking performance and output-format reliability during SFT. Parse failures receive a score of zero in the NDCG calculation.}
    \label{fig:main-training-dynamics}
\end{figure}

Figure~\ref{fig:main-training-dynamics} shows all-sample NDCG and parse-failure rate throughout training. SCOReD maintains higher all-sample NDCG while producing substantially fewer malformed outputs. Although the parse-failure rate of the LLM-summarization baseline decreases during training, it remains considerably higher than those of the other methods. These results show that minimizing trace length alone does not produce an effective distillation target. Aggressive one-shot compression can remove useful reasoning or weaken adherence to the required output format.

\subsection{Post-SFT Optimization}
\label{sec:post-sft-results}

We further investigate whether on-policy optimization, such as Reinforcement Learning with DAPO \citep{yu2026dapo} and On-Policy Self Distillation (OPSD) \citep{opsd}, can improve the SCOReD SFT student.

\paragraph{On-Policy Self-Distillation} Standard On-Policy Distillation (OPD) \citep{opd} requires a teacher that is both more capable than the student and compatible with the student's vocabulary. Gemma-4-26B provides a stronger task policy but does not share the Qwen student's vocabulary. Qwen-3.6-35B-A3B is vocabulary-compatible with the student but does not provide a consistently stronger policy on this task. We therefore evaluate On-policy Self-Distillation with privileged information (OPSD) instead of standard on-policy distillation. Concretely, we include the ground truth in the prompt itself as privileged information. Since the ground truth is a selection rather thank ranking of all candidates, we take the teacher's ranking and permute the ranking such that the ground truth items are ranked first.

We find that OPSD does not improve over the SFT initialization. On the complete test set, the initial SFT checkpoint obtains the best NDCG, MRR, and MAP. OPSD checkpoints exhibit small, non-monotonic fluctuations, and do not show a general improvement in ranking quality. The same pattern holds when evaluation is restricted to a fixed correctly parsed subset. Detailed OPSD results are reported in Appendix~\ref{app:opsd-results}.

\paragraph{Reinforcement Learning with DAPO} Direct reinforcement learning with DAPO similarly fails to improve the SFT model. NDCG remains within a narrow range over six epochs, while parse failures fluctuate without a sustained downward trend. Because the SFT student already approaches the performance of its Gemma-4-26B teacher, the remaining improvements available to policy optimization are small. This behavior is consistent with a sparse or low-variance reward signal, in which sampled rankings often receive similar rewards and therefore provide limited positive advantage. The complete DAPO trajectory is shown in Appendix~\ref{app:dapo-results}.

\section{Conclusion}
\label{sec:conclusion}

We introduced SCOReD, a student-aware framework for compressing recommendation reasoning traces through LLM-based segmentation, attention-based importance scoring, and reward-guided edit selection. Our analysis shows that teacher traces contain substantial redundancy, particularly in repeated verification and ranking stages that rarely alter the final prediction.

Training on SCOReD-optimized traces improves NDCG while reducing average trace length by approximately $27\%$ and parse failures by approximately $46\%$ relative to standard SFT. The resulting $0.6$B student also outperforms Qwen-3.6-35B-A3B on most reported metrics. In contrast, one-shot LLM summarization compresses more aggressively but substantially reduces output reliability, while OPSD and DAPO provide no consistent improvement over the SCOReD SFT model. These results demonstrate that effective CoT compression should be guided by the behavior of the student model rather than by length reduction alone.

\bibliographystyle{assets/plainnat}
\bibliography{paper}

\clearpage
\newpage
\beginappendix

\section{CoT Segmentation}
\label{app:cot_segment_prompt}

\begin{promptbox}{Segmentation System Prompt}{blue}
You are an expert data annotator. Your task is to segment a reranking reasoning chain into contiguous blocks belonging to one of the following stages:
\medskip
H: Purchase History Analysis (summarizing or categorizing past purchases)

P: User Preference / Theme Modeling (inferring categories, brands, ingredients, benefits, or use cases from history)

C: Candidate Analysis (describing candidate items and their relevance before constructing the ranking)

R: Ranking Construction / Tiering (building tiers, assigning rank positions, or forming a preliminary ordered list)

V: Verification / Pairwise Comparison / Self-Correction (checking close calls, comparing candidates, expressing doubt, or revising the order)

F: Final Ranking (the final conclusion, final order, or final ranked list inside the reasoning chain)

Common reranking patterns include candidate summaries, high/mid/low tiers, pairwise comparisons like 'candidate 5 vs candidate 6', repeated 'Wait' checks, and one or more final ordered lists.
\medskip
\textbf{CRITICAL RULES}:
1. You will be given the reasoning chain with line numbers.
2. Output your response STRICTLY as a JSON array of objects. Each object must have 'stage' (string, one of H, P, C, R, V, F), 'start\_line' (int), and 'end\_line' (int).
3. The segments must cover all lines from {first\_line} to {final\_line} continuously, with no gaps or overlaps.
4. Segment at HIGH GRANULARITY. Break down long blocks when the reasoning shifts between history, preferences, candidate analysis, tiering, verification, and final ranking.
5. Do not label every ordered list as F. Use F only for final conclusions or final ranking statements; use R for intermediate tiering and draft rankings.
6. Output ONLY the valid JSON array. Do not include markdown formatting like \textasciigrave{}\textasciigrave{}\textasciigrave{}json.
\end{promptbox}

\begin{promptbox}{User Turn Input}{violet}

*   Purchase 1: 100\% Pure Unrefined Raw Shea Butter (1 Pound). Focus: Unrefined, Raw, Shea Butter, natural, skin and hair moisturizing/nourishing.

*   Purchase 2: Raw African Black Soap Imported From Ghana (1lb 16oz). Focus: African Black Soap, natural, hand-made, West Africa, moisturizing, skin cleanser.
\medskip
*   Candidate 1: Raw African Shea Butter Black Soap from Ghana - 1 Lb. (Shea Butter + Black Soap. Matches both previous purchases perfectly).
*   Candidate 2: MDSolarSciences - SPF 40 Mineral Screen Lotion. (Sunscreen. Skincare, but a different category - sun protection).
*   Candidate 3: Raw African Black Soap from Ghana 1 Lb. (Pure Black Soap. Matches Purchase 2 perfectly).
*   Candidate 4: Yes To Tomatoes Skin Clearing Facial Mask. (Skincare, facial mask. Less direct connection than Shea/Black Soap).
*   Candidate 5: Earth Therapeutics: Anti-Stress Microwavable Neck Pillow. (Wellness/relaxation, no direct connection to Shea/Black Soap products).
*   Candidate 6: Tate's The Natural Miracle - Odorless Nail Polish Remover. (Nail care, not highly relevant).
*   Candidate 7: TRESemme Climate Control Conditioner. (Hair care, but mass-market/commercial, whereas user prefers raw/unrefined).
*   Candidate 8: e.l.f. Cosmetics Eyelash \& Brow Wand. (Makeup, different category).
*   Candidate 9: 100g Pure USAWHR Indigo Powder. (Natural hair dye. Relevant to the "natural hair care" aspect).
*   Candidate 10: Seaextend Ultimate Colorcare Silkening Oil Treatment. (Hair styling treatment/oil. Somewhat related to hair care).
\medskip
*   \textbf{High Match:}
        *   Candidate 1 (Shea Butter + Black Soap): Combines both interests.
        *   Candidate 3 (Black Soap): Direct match to Purchase 2.
*   \textbf{Medium Match:}
        *   Candidate 9 (Indigo Powder): Natural hair care/dye, aligns with the "natural/pure" theme of Purchase 1 \& 2.
        *   Candidate 10 (Silkening Oil Treatment): Hair care, but less "raw/natural" feeling.
        *   Candidate 4 (Facial Mask): Skincare, but different type.
        *   Candidate 2 (Mineral Sunscreen): Skincare, but functional (sun protection).
*   \textbf{*Low Match:*}
        *   Candidate 7 (TRESemme): Hair care, but mass-market (contradicts the unrefined/pure preference).
        *   Candidate 8 (Eyelash/Brow Wand): Makeup.
        *   Candidate 6 (Nail Polish Remover): Nail care.
        *   Candidate 5 (Neck Pillow): Wellness, but unrelated.
\medskip
    1.  Candidate 1 (Shea + Black Soap) - Best fit.
    2.  Candidate 3 (Black Soap) - Very strong fit.
    3.  Candidate 9 (Indigo Hair Dye) - Natural hair care/pure/no-chemical aspect matches the "unrefined" preference.
    4.  Candidate 10 (Hair Silkening Oil) - Hair care.
    5.  Candidate 4 (Facial Mask) - Skincare.
    6.  Candidate 2 (Mineral Sunscreen) - Skincare.
    7.  Candidate 7 (TRESemme Conditioner) - Hair care (though mass market).
    8.  Candidate 8 (Eyelash/Brow Wand) - Makeup.
    9.  Candidate 6 (Nail Polish Remover) - Nail care.
    10. Candidate 5 (Neck Pillow) - Wellness/Relaxation (User mentioned "relaxes" in Shea description, but this is a stretch).
\medskip
    *   User likes: Raw, Unrefined, Shea Butter, African Black Soap, natural ingredients.
    *   C1: Shea Butter + Black Soap (Excellent match).
    *   C3: Black Soap (Excellent match).
    *   C9: Natural Indigo hair dye (Matches "natural/pure/unrefined" theme for hair/skin).
    *   C10: Silkening oil (Hair focus).
    *   C4: Facial mask (Skincare focus).
    *   C2: Sunscreen (Skincare focus).
    *   C7: Conditioner (Hair care focus).
    *   C8: Brow wand (Makeup).
    *   C6: Nail polish remover (Nail care).
    *   C5: Neck Pillow (Wellness).
\medskip
\textbf{Refining Ranking:}
    C1 (1)
    C3 (3)
    C9 (9)
    C10 (10)
    C4 (4)
    C2 (2)
    C7 (7)
    C8 (8)
    C6 (6)
    C5 (5)
\medskip
    Wait, let's re-evaluate C7 vs C4/C2. C7 is hair care (user uses Shea for hair), but it's TRESemme (commercial/processed). User emphasized "Unrefined" and "nothing wrong... as long as it is Unrefined... when it's not listed as RAW... it is not worth buying." This heavily weights against C7 and in favor of C9.
\medskip
    C1 and C3 are definitely 1 and 2.
    C9 is 3 because it's pure/natural (100\% Indigo).
    C10 is 4 (Hair care, though maybe not "natural" enough).
    C4 is 5 (Skincare).
    C2 is 6 (Skincare).
    C7 is 7 (Hair care, commercial).
    C8 is 8 (Makeup).
    C6 is 9 (Nail care).
    C5 is 10 (Wellness).
\medskip
    1: Shea/Soap
    3: Soap
    9: Indigo (Natural hair)
    10: Silkening Oil (Hair care)
    4: Mask (Skincare)
    2: Sunscreen (Skincare)
    7: Conditioner (Hair care)
    8: Brow Wand (Makeup)
    6: Nail Polish (Nail care)
    5: Neck Pillow (Wellness)
\medskip
    Wait, C3 is "Raw African Black Soap" and C1 is "Raw African Shea Butter Black Soap". The user bought Shea Butter and Black Soap separately. C1 combines them. C3 is just Black Soap.
    C9 (Indigo) is very thematic with "Unrefined/Pure/Natural/No Chemicals".
\medskip
    Double check C1 and C3.
    C1: Raw African Shea Butter Black Soap from Ghana - 1 Lb.
    C3: Raw African Black Soap from Ghana 1 Lb.
    C1 includes shea butter, which the user also bought.
\medskip
    \textbf{Final list Check:} [1, 3, 9, 10, 4, 2, 7, 8, 6, 5]
\medskip
    \textbf{Let's quickly re-check the "relaxes" part of Shea butter:} Shea is mentioned to help "relax" the body/hair. Candidate 5 (Neck pillow) mentions "relaxing" and "soothing", but it's far removed from the skin/hair/raw products theme.
\medskip
    \textbf{Is there any other candidate closer to Shea?}
    Candidate 10 is hair oil.
    Candidate 7 is conditioner.
    Candidate 9 is natural dye.
\medskip
    Ranking looks solid.
\medskip
    1. Candidate 1 (Shea + Soap)
    2. Candidate 3 (Soap)
    3. Candidate 9 (Natural Indigo)
    4. Candidate 10 (Hair Oil)
    5. Candidate 4 (Facial Mask)
    6. Candidate 2 (Sunscreen)
    7. Candidate 7 (Conditioner)
    8. Candidate 8 (Eyelash Wand)
    9. Candidate 6 (Nail Polish Remover)
    10. Candidate 5 (Neck Pillow)
\end{promptbox}

\begin{jsonblock}{Assistant Turn Output}{teal}
[
    {"stage": "H", "start_line": 0, "end_line": 2},
    {"stage": "C", "start_line": 3, "end_line": 13},
    {"stage": "R", "start_line": 14, "end_line": 38},
    {"stage": "V", "start_line": 39, "end_line": 44},
    {"stage": "C", "start_line": 45, "end_line": 50},
    {"stage": "R", "start_line": 51, "end_line": 62},
    {"stage": "V", "start_line": 63, "end_line": 74},
    {"stage": "R", "start_line": 75, "end_line": 84},
    {"stage": "H", "start_line": 85, "end_line": 85},
    {"stage": "V", "start_line": 86, "end_line": 104},
    {"stage": "F", "start_line": 105, "end_line": 114}
]
\end{jsonblock}
\section{Qualitative Comparison of Compression Algorithms}

This section showcases the result of compression on a sample CoT generated by Gemma-4-26B. The prompt contains six items in the user's purchase history.

\begin{promptbox}{User Prompt}{violet}

\textbf{User's purchase history:}
Purchase 1
Title: Motions Naturally You, Hydrate My Curls Pudding 8 Ounce
Description: [Missing]
--------

Purchase 2
Title: Envyderm Eyelash Enhancement and Conditioning Nighttime Serum, 5ml (0.17oz)
Description: EnvyDerm Eyelash Enhancement; Conditioning Nighttime Serum uses a specially formulated blend of ingredients to promote your natural keratin production in the lash line to help the follicle grow naturally stronger lashes. EnvyDerm's latest Peptides technology is clinically proven to increase length and density of lashes by 25\% and by as much as 72\% in 6 weeks of use. In addition, Proteins, Vitamins and Moisturizing Oils nourish, lengthen and strengthen the eyelash hairs. Our product contains only safe natural ingredients that have been proven to grow and thicken eyelashes effectively. The serum has undergone clinical tests to prove effectiveness The Serum uses the leading peptide technology to induce keratin production in the lash line thus promoting longer, stronger enhanced lashes. Moreover the proteins, vitamins, minerals and moisturizing oils fortify, nourish and add shine to for beautiful enhanced lashes and brows. Lashes lose fullness due to everyday wear and tear, make-up removal and the aging process. Eyelash hairs are quicker to break, more prone to fall out, and slower to grow than other hair. This is why it's so important to defend the eyelashes. EnvyDerm Eyelash Enhancement and Conditioning Nighttime Serum should last from 3 to 6 months depending on personal use. Expect three months of use on the upper and lower lash lines and brows and 6 months if only used on the upper lash line.
--------

Purchase 3
Title: Hydroxatone Intensive Anti-Wrinkle Complex
Description: [Missing]
--------

Purchase 4
Title: Tresemme Keratin Smooth Infusing Shampoo, 25 Ounce
Description: [Missing]
--------

Purchase 5
Title: Schick Hydro 5 Disposable Razor, 3 Count
Description: [Missing]
--------

Purchase 6
Title: Schick Hydro Silk Disposable Razor, 3 Count
Description: [Missing]
--------

\textbf{Candidate items:}
Candidate 1
Title: Organic Root Stimulator Olive Oil Replenishing Conditioner, 12.25 Ounce
Description: Deep penetrating conditioner. Helps restore moisture and rebuild damaged hair. Stimulates the root of your hair. It leaves hair moisturized, soft, and healthy.
--------

Candidate 2
Title: China Glaze Crackle Collection Lightening Bolt [Misc.]
Description: China Glaze Lightning Bolt is a white crackle nail polish color. China Glaze Crackle Collection, Spring 2011.
--------

Candidate 3
Title: Tweezerman LTD Deluxe Metal Eyelash Curler
Description: Our best-selling eyelash curler is expertly crafted to create  beautifully enduring curl.
--------

Candidate 4
Title: L'Oreal Paris Age Perfect Hydra-Nutrition Moisturizer, 1.7-Fluid Ounce
Description: Mature, very dry skin needs extra care and attention. Treat it to deep hydration with Age Perfect Hydra-Nutrition Anti-Sagging Ultra-Nourishing Day/Night Cream.  Skin is nourished with a nutrient complex enriched with calcium that builds resilience, leaving the skin feeling fortified while alleviating uncomfortable dryness.  This creamy and silky texture melts away leaving skin feeling soft and nourished.  Skin is instantly more hydrated, and in 2 weeks comfort is restored to the skin.  After 4 weeks skin is firmer and more supple.
--------

Candidate 5
Title: Burt's Bees Hair Repair Shea; Grapefruit Deep Conditioner, 5 Fluid Ounces
Description: Burt's Bees Grapefruit Conditioner moisturizes dry hair with Natural Shea Butter and Coconut Oil, while Grapefruit Extract and essential Citrus Oils help to add shine. SLS-free and designed for especially dry hair, use this deep conditioner weekly to moisturize and nourish for silky, soft, and beautiful hair.
--------

Candidate 6
Title: Cristophe Professional Glossing Shampoo, 10 Ounce
Description: Cristophe Professional Glossing Shampoo provides instant shine and UV protection for luminous hair and helps you create soft, silky styles just like Cristophe. Exotic Argan and Monoi oils combine to soften hair and protect strands from styling damage and breakage. This lightweight salon formula smoothes the cuticle for glossy, shiny hair.
--------

Candidate 7
Title: Essie Nail Lacquer, Midnight Cami, 0.5 Fluid Ounce
Description: Essie nail lacquer goes on smooth and last long Deep blue shade Chip-resistant DBP, Toluene and Formaldehyde free.
--------

Candidate 8
Title: Remington S8500 Shine Therapy  Moisturizing and Conditioning Digital Ceramic Hair Straightener, 1\&quot;
Description: [Missing]
--------

Candidate 9
Title: Manic Panic Pillarbox Red 4oz
Description: Manic Panic presents semi-permanent hair dye, with a longer-lasting color! For best results, pre-lighten hair. Use latex gloves. Made in the USA.
--------

Candidate 10
Title: Dermalogica PreCleanse 5.1 oz.
Description: A lightweight Olive and Apricot Kernel-based cleansing oil, fortified with Kukui Nut, Borage Seed and Rice Bran oils, that gently liquefies sebum and oil-based debris from the skin's surface. The hydrophilic (water-loving) formula enables PreCleanse and oil-based debris to be readily rinsed away without leaving behind any greasy residue. Follow with prescribed Dermalogica Cleanser for a thorough double cleansing. Contains no artificial fragrance or color.
--------    
\end{promptbox}

\paragraph{Teacher's Raw CoT:} Gemma-4-26B's raw CoT shows substantial uncertainty, as it revisits the ranking eight times. However, only the first two revisions change the ordering. The initial ranking was (\texttt{[1,5,6,4,3,8,10,9,2,7]}), which prioritized Candidates (1), (5), and (6) because they all matched the user's preference for moisturizing and smoothing hair-care products. In the first revision, Candidate (6) was moved to the top because it is a shampoo and therefore provides the closest product-type match to the user's previously purchased TRESemmé shampoo. Candidate (8) was also temporarily placed above Candidate (3), since the teacher viewed the hair straightener's smoothing function as closely related to the user's keratin-smoothing hair product. This produced (\texttt{[6,1,5,4,8,3,10,9,2,7]}).

In the second revision, Candidate (3) was moved back above Candidate (8), as the eyelash curler was considered a more direct category match to the user's eyelash serum. The resulting ranking, (\texttt{[6,1,5,4,3,8,10,9,2,7]}), remained unchanged. The final six verification rounds repeatedly reconsidered the same comparisons, particularly Candidates (1), (5), and (6), and Candidates (3) and (8), without introducing new evidence. This suggests that the repeated checking was caused by uncertainty over whether product type, category, or functional similarity should be prioritized, as well as the absence of a clear stopping criterion.

\begin{promptbox}{Teacher's Raw CoT}{BrickRed}
<think>
\textcolor{gray}{$\triangleright$ Purchase History Analysis}
*   Purchase 1: Motions Naturally You, Hydrate My Curls Pudding (Hair care, curly hair, moisturizing)
    *   Purchase 2: Envyderm Eyelash Enhancement Serum (Eye/lash care, nighttime serum)
    *   Purchase 3: Hydroxatone Intensive Anti-Wrinkle Complex (Anti-aging skin care)
    *   Purchase 4: Tresemme Keratin Smooth Infusing Shampoo (Hair care, keratin/smoothing)
    *   Purchase 5: Schick Hydro 5 Disposable Razor (Shaving/Grooming)
    *   Purchase 6: Schick Hydro Silk Disposable Razor (Shaving/Grooming)
    
\medskip
\textcolor{gray}{$\triangleright$ User Preference / Theme Modeling}
*Key themes: Hair Care (moisturizing, smoothing, keratin), Skin Care (anti-aging, moisturizing), Eye Care (lashes), Grooming (shaving).*

\medskip
\textcolor{gray}{$\triangleright$ Candidate Analysis}
*   Title: Organic Root Stimulator Olive Oil Replenishing Conditioner
        *   Category: Hair Care (Conditioner, moisturizing/restorative)
        *   Match level: High (Hair care, similar to P1, P4).

*   Title: China Glaze Crackle Collection Lightening Bolt
        *   Category: Nail Care (Nail polish)
        *   Match level: Low (No nail care in history).

*   Title: Tweezerman LTD Deluxe Metal Eyelash Curler
        *   Category: Eye/Lash Care (Tool)
        *   Match level: Medium-High (Eye care connection to P2).

*   Title: L'Oreal Paris Age Perfect Hydra-Nutrition Moisturizer
        *   Category: Skin Care (Anti-aging, moisturizing)
        *   Match level: High (Skin care, similar to P3).

*   Title: Burt's Bees Hair Repair Shea \& Grapefruit Deep Conditioner
        *   Category: Hair Care (Conditioner, moisturizing)
        *   Match level: High (Hair care, similar to P1, P4).

*   Title: Cristophe Professional Glossing Shampoo
        *   Category: Hair Care (Shampoo, smoothing/shine)
        *   Match level: High (Hair care, similar to P4).

*   Title: Essie Nail Lacquer, Midnight Cami
        *   Category: Nail Care (Nail polish)
        *   Match level: Low (No nail care in history).

*   Title: Remington S8500 Shine Therapy Hair Straightener
        *   Category: Hair Care (Tool, smoothing)
        *   Match level: High (Hair care connection to P4, P1).

*   Title: Manic Panic Pillarbox Red
        *   Category: Hair Care (Hair dye)
        *   Match level: Medium (Hair care, but dye is different from smoothing/hydrating).

*   Title: Dermalogica PreCleanse
        *   Category: Skin Care (Cleansing oil)
        *   Match level: Medium-High (Skin care, similar to P3).

\medskip
\textcolor{gray}{$\triangleright$ Verification / Pairwise Comparison / Self-Correction}

1.  Candidates 1, 4, 5, 6 (Directly matching primary categories: Hair Care and Skin Care).
        2.  Candidate 3 (Eye care - specifically related to P2).
        3.  Candidate 8 (Hair styling tool - related to hair care).
        4.  Candidate 10 (Skin care - related to P3).
        5.  Candidate 9 (Hair care - dye, slightly less relevant than smoothing/hydrating).
        6.  Candidates 2, 7 (Nail care - not in history).


\medskip
\textcolor{gray}{$\triangleright$ Verification / Pairwise Comparison / Self-Correction}

*   P1 (Hydrate Curls) + P4 (Keratin Smooth Shampoo) -> Candidates 1, 5, 6 (Conditioners/Shampoo).
    *   P3 (Anti-wrinkle Complex) -> Candidate 4 (Hydrating moisturizer).
    *   P2 (Eyelash Serum) -> Candidate 3 (Eyelash Curler).
    *   P5/P6 (Razors) -> No specific match, but grooming is the category.

\medskip
\textcolor{gray}{$\triangleright$ Verification / Pairwise Comparison / Self-Correction}

*Re-evaluating Candidates 1, 5, 6 (Hair Care):*
    *   Candidate 1 (Conditioner): Very high match for hair maintenance.
    *   Candidate 5 (Conditioner): Very high match for hair maintenance.
    *   Candidate 6 (Shampoo): Very high match.

    *Re-evaluating Candidate 4 (Skin care):*
    *   Candidate 4 (Anti-aging moisturizer): Very high match for P3 (Anti-wrinkle).

    *Comparing Candidate 1 vs 5 vs 6 vs 4:*
    *   The user has two hair products (P1, P4), one skin product (P3), one lash product (P2).
    *   Hair care is the most frequent category.
    *   Let's prioritize hair care, then skin care, then lash care.

\medskip
\textcolor{gray}{$\triangleright$ Ranking Construction / Tiering}

*Refined ranking order:*
    1.  Candidate 1 (Hair Care - Conditioner)
    2.  Candidate 5 (Hair Care - Conditioner)
    3.  Candidate 6 (Hair Care - Shampoo)
    4.  Candidate 4 (Skin Care - Anti-aging moisturizer)
    5.  Candidate 3 (Eye Care - Eyelash curler)
    6.  Candidate 8 (Hair Care - Tool)
    7.  Candidate 10 (Skin Care - Cleanser)
    8.  Candidate 9 (Hair Care - Dye)
    9.  Candidate 2 (Nail Care)
    10. Candidate 7 (Nail Care)

\medskip
\textcolor{gray}{$\triangleright$ Verification / Pairwise Comparison / Self-Correction}

*Let's look closer at themes:*
    *   P1 \& P4 are about *moisture* and *smoothing* hair.
    *   Candidate 1: "helps restore moisture and rebuild damaged hair". (Strong)
    *   Candidate 5: "moisturizes dry hair... for silky, soft, and beautiful hair". (Strong)
    *   Candidate 6: "provides instant shine... smoothes the cuticle". (Strong - very similar to Tresemme Keratin Smooth).
    *   Candidate 4: "mature, very dry skin... anti-sagging... anti-aging". (Strong match for P3 Hydroxatone).
    *   Candidate 3: "eyelash curler". (Connection to P2).
    *   Candidate 8: "moisturizing and conditioning hair straightener". (Connection to P4).
    *   Candidate 10: "cleansing oil... facial cleanser". (Connection to skin care).
    *   Candidate 9: "hair dye". (Hair care but distinct category).
    *   Candidate 2/7: Nail polish (No match).

\medskip
\textcolor{gray}{$\triangleright$ Ranking Construction / Tiering}


*Rank refinement:*
    1.  Candidate 6 (Shampoo, very close to P4 Tresemme Keratin Smooth)
    2.  Candidate 1 (Conditioner, matches theme of hydrating/rebuilding)
    3.  Candidate 5 (Conditioner, matches theme of hydrating/moisturizing)
    4.  Candidate 4 (Anti-aging skin cream, directly relates to P3)
    5.  Candidate 8 (Hair straightener, relates to smoothing P4)
    6.  Candidate 3 (Eyelash curler, relates to P2)
    7.  Candidate 10 (Cleansing oil, relates to skin care)
    8.  Candidate 9 (Hair dye, hair care)
    9.  Candidate 2 (Nail polish)
    10. Candidate 7 (Nail polish)

\medskip
\textcolor{gray}{$\triangleright$ Verification / Pairwise Comparison / Self-Correction}

*Self-Correction:*
    - Candidate 6 is "Glossing Shampoo". P4 is "Keratin Smooth Shampoo".
    - Candidate 1 is "Olive Oil Replenishing Conditioner".
    - Candidate 5 is "Shea \& Grapefruit Deep Conditioner".
    - Both 1 and 5 are conditioners. Let's look at P1: "Hydrate My Curls Pudding". Pudding is a styling/moisture product.
    - Candidate 4 is "Age Perfect Hydra-Nutrition Moisturizer". Very strong connection to P3 "Hydroxatone Intensive Anti-Wrinkle Complex".

    *Let's re-order:*
    Hair Care (1, 5, 6) > Skin Care (4, 10) > Lash Care (3) > Hair Tool (8) > Hair Dye (9) > Nail Care (2, 7).
    Wait, P4 is a shampoo. Candidate 6 is a shampoo.
    P1 is a hydrating hair product. Candidate 1/5 are conditioners.
    P3 is an anti-wrinkle product. Candidate 4 is an anti-aging/hydrating product.
    P2 is an eyelash product. Candidate 3 is an eyelash product.


\medskip
\textcolor{gray}{$\triangleright$ Ranking Construction / Tiering}

*Revised ranking order attempt:*
    1.  Candidate 6 (Shampoo, direct type match to P4)
    2.  Candidate 1 (Conditioner, theme of moisture match to P1, P4)
    3.  Candidate 5 (Conditioner, theme of moisture match to P1, P4)
    4.  Candidate 4 (Anti-aging moisturizer, direct theme match to P3)
    5.  Candidate 3 (Eyelash tool, direct theme match to P2)
    6.  Candidate 8 (Hair tool, theme match to P4)
    7.  Candidate 10 (Skin cleanser, theme match to skin care)
    8.  Candidate 9 (Hair dye, hair care)
    9.  Candidate 2 (Nail polish)
    10. Candidate 7 (Nail polish)

\medskip
\textcolor{gray}{$\triangleright$ Verification / Pairwise Comparison / Self-Correction}

*Let's consider if Candidate 8 (hair straightener) should be higher:* It's a hair tool. The user bought hair care products, not tools. Candidate 3 (eyelash curler) is a tool for an eye product they bought.
    Actually, Candidate 4 (Skin) is extremely similar to P3 (Skin).
    
\medskip
\textcolor{gray}{$\triangleright$ Ranking Construction / Tiering}

*Finalizing the order:*
    [6, 1, 5, 4, 3, 8, 10, 9, 2, 7]

\medskip
\textcolor{gray}{$\triangleright$ Purchase History Analysis}

*Checking hair care products (1, 4, 5, 6, 8, 9) vs skin care (3, 4, 10) vs lash (2, 3) vs nail (2, 7):*
    *P1: Hair (Curl/Hydrate)
    *P2: Lash (Serum/Growth)
    *P3: Skin (Anti-wrinkle)
    *P4: Hair (Shampoo/Smooth)
    *P5: Grooming (Razor)
    *P6: Grooming (Razor)
    
\medskip
\textcolor{gray}{$\triangleright$ Candidate Analysis}

Candidate 1: Hair (Conditioner)
    Candidate 2: Nail
    Candidate 3: Lash (Curler)
    Candidate 4: Skin (Anti-aging)
    Candidate 5: Hair (Conditioner)
    Candidate 6: Hair (Shampoo)
    Candidate 7: Nail
    Candidate 8: Hair (Tool/Straightener)
    Candidate 9: Hair (Color)
    Candidate 10: Skin (Cleanser)

\medskip
\textcolor{gray}{$\triangleright$ Ranking Construction / Tiering}

*Ranking:*
    1. Candidate 6 (Shampoo - P4)
    2. Candidate 1 (Conditioner - P1, P4)
    3. Candidate 5 (Conditioner - P1, P4)
    4. Candidate 4 (Anti-aging - P3)
    5. Candidate 3 (Lash tool - P2)
    6. Candidate 8 (Hair tool - P4)
    7. Candidate 10 (Skin cleanser - P3)
    8. Candidate 9 (Hair dye - P1, P4)
    9. Candidate 2 (Nail)
    10. Candidate 7 (Nail)

\medskip
\textcolor{gray}{$\triangleright$ Verification / Pairwise Comparison / Self-Correction}

*Wait, let's look at Candidate 4 and 10:*
    Candidate 4 is "Age Perfect Hydra-Nutrition Moisturizer" (Anti-aging skin).
    Candidate 10 is "Dermalogica PreCleanse" (Cleansing oil).
    P3 is "Hydroxatone Intensive Anti-Wrinkle Complex".
    4 is a much better match than 10.

    *Look at 1 and 5 again:*
    Candidate 1: Organic Root Stimulator Olive Oil Replenishing Conditioner
    Candidate 5: Burt's Bees Hair Repair Shea \& Grapefruit Deep Conditioner
    P1: Motions Naturally You, Hydrate My Curls Pudding.
    P4: Tresemme Keratin Smooth Infusing Shampoo.
    1 and 5 are both solid hair conditioners.

    *Look at 8 and 3 (Tools):*
    Candidate 3: Eyelash curler (relates to P2).
    Candidate 8: Hair straightener (relates to P4).
    The user bought lashes serum (P2) and shampoo/hair products. 3 and 8 are tools.

\medskip
\textcolor{gray}{$\triangleright$ Ranking Construction / Tiering}

*Final Sequence check:*
    6 (Shampoo)
    1 (Conditioner)
    5 (Conditioner)
    4 (Anti-aging skin)
    3 (Eyelash tool)
    8 (Hair tool)
    10 (Skin cleanser)
    9 (Hair dye)
    2 (Nail)
    7 (Nail)

\medskip
\textcolor{gray}{$\triangleright$ Verification / Pairwise Comparison / Self-Correction}

Actually, let's compare 3 and 8 again. Candidate 3 is an eyelash tool. The user bought eyelash *serum*. Candidate 8 is a hair *straightener*. The user bought *shampoo* and *hair pudding*.
    Candidate 1, 5, 6 (Hair care) are definitely top.

Candidate 4 (Skin care) is next.
    Then the tools (3, 8).
    Then the rest (10, 9, 2, 7).

\medskip
\textcolor{gray}{$\triangleright$ Verification / Pairwise Comparison / Self-Correction}

*Is 6, 1, or 5 better?*
    P4 is Tresemme Keratin Smooth Shampoo. Candidate 6 is Cristophe Glossing Shampoo.
    P1 is Motions Hydrate My Curls Pudding. Candidate 1 is Root Stimulator Conditioner.
    I'll keep 6 as 1st.

    Wait, let's check if 8 should be higher. P4 is about "Keratin Smooth". A straightener (8) helps achieve that. But products are generally more preferred if you're already buying them.

\medskip
\textcolor{gray}{$\triangleright$ Verification / Pairwise Comparison / Self-Correction}

One last check:
    P1, P4 (Hair care)
    P2 (Lash care)
    P3 (Skin care)
    P5, P6 (Grooming)

    Candidate 1: Hair
    Candidate 2: Nail
    Candidate 3: Lash
    Candidate 4: Skin
    Candidate 5: Hair
    Candidate 6: Hair
    Candidate 7: Nail
    Candidate 8: Hair tool
    Candidate 9: Hair dye
    Candidate 10: Skin

\medskip
\textcolor{gray}{$\triangleright$ Ranking Construction / Tiering}

Rank:
    6, 1, 5 (Hair products)
    4 (Skin product)
    3 (Lash tool)
    8 (Hair tool)
    10 (Skin cleanser)
    9 (Hair dye)
    2, 7 (Nail)

\medskip
\textcolor{gray}{$\triangleright$ Verification / Pairwise Comparison / Self-Correction}

Let's try to be more precise with 6, 1, 5.
    P4 is Tresemme. Candidate 6 is Cristophe.
    P1 is Motions. Candidate 1 is Organic Root.
    Actually, Candidate 5 (Burt's Bees) might be more similar to P1's moisturizing focus? "Burt's Bees... moisturizes dry hair... Shea \& Grapefruit". P1 is "Hydrate My Curls". Let's put 5 or 1 near 6.

\medskip
\textcolor{gray}{$\triangleright$ Verification / Pairwise Comparison / Self-Correction}

Most important are the categories.

    1. Hair care (6, 1, 5, 9, 8)
    2. Skin care (4, 10)
    3. Lash care (3)
    4. Nail care (2, 7)

    Let's re-rank the Hair Care:
    P4: Keratin Smooth Shampoo.
    Candidate 6: Glossing Shampoo. (Very close)
    P1: Hydrate Curls Pudding.
    Candidate 1: Replenishing Conditioner.
    Candidate 5: Deep Conditioner.
    Candidate 9: Hair dye (different).
    Candidate 8: Straightener (tool).

\medskip
\textcolor{gray}{$\triangleright$ Ranking Construction / Tiering}

Ranking within hair: 6, 1, 5, 8, 9.
    Wait, Candidate 9 (Hair dye) is a product, whereas 8 is a tool. Usually, products are prioritized. But 8 is a hair tool and the user bought hair products.

    Re-evaluating the whole list:
    6 (Hair Shampoo)
    1 (Hair Conditioner)
    5 (Hair Conditioner)
    4 (Skin Moisturizer - P3)
    3 (Lash tool - P2)
    8 (Hair tool - P4)
    10 (Skin Cleanser - P3)
    9 (Hair Dye - P1/4)
    2 (Nail Polish)
    7 (Nail Polish)

\medskip
\textcolor{gray}{$\triangleright$ Verification / Pairwise Comparison / Self-Correction}

*Wait:* Let's re-examine Candidate 3 vs Candidate 8.
    P2 is lash serum. Candidate 3 is eyelash curler. (Direct category)
    P4 is smoothing shampoo. Candidate 8 is smoothing straightener. (Direct theme)
    This is a toss-up. I'll put 3 first since it's a direct category match (eye care).

    *Wait:* Let's re-examine 1, 5, 6 again.
    6: Shampoo (P4 is Shampoo)
    1: Conditioner
    5: Conditioner
    This seems most logical.

    *Double check Candidate 4 (Skin) vs Candidate 3 (Lash):*
    P3 (Skin) and P2 (Lash).
    Candidate 4 (Skin) and Candidate 3 (Lash).
    4 is a strong skin match. 3 is a lash tool. Skin care is generally more consistent with the user's P3 than 3 is with P2.

    *Final check on Candidate 9 (Dye):* Dye is a hair product, but it's a very different sub-category from "Hydrate" or "Smooth".


\medskip
\textcolor{gray}{$\triangleright$ Ranking Construction / Tiering}

*Final list construction:*
    6 (Shampoo)
    1 (Conditioner)
    5 (Conditioner)
    4 (Skin moisturizer)
    3 (Lash tool)
    8 (Hair tool)
    10 (Skin cleanser)
    9 (Hair dye)
    2 (Nail polish)
    7 (Nail polish)

\medskip
\textcolor{gray}{$\triangleright$ Verification / Pairwise Comparison / Self-Correction}

(Wait, candidate 1 and 5 are quite similar, 6 is the only shampoo. Let me ensure 6, 1, 5 or 6, 5, 1).
    Candidate 1: Organic Root Stimulator Olive Oil Replenishing Conditioner (12.25 oz)
    Candidate 5: Burt's Bees Hair Repair Shea \& Grapefruit Deep Conditioner (5 oz)
    Both are conditioners. 6 (shampoo) is the closest match to P4 (shampoo).

    Let's look at the order again.
    [6, 1, 5, 4, 3, 8, 10, 9, 2, 7]
    One more check: 4 (Skin) vs 3 (Lash).
    P3 is an anti-wrinkle complex. Candidate 4 is Age Perfect Hydra-Nutrition Moisturizer.
    P2 is an eyelash serum. Candidate 3 is an eyelash curler.
    Candidate 4 is a more direct "use-case" match to P3.
</think>

[6, 1, 5, 4, 3, 8, 10, 9, 2, 7]    
\end{promptbox}


\paragraph{LLM Summarized CoT:} The LLM-compressed CoT preserves the meaningful refinement stages from the teacher's original trace. In particular, it retains the decision to prioritize Candidate (6) because of its direct shampoo match, the comparison between Candidates (3) and (8), and the distinction between the stronger skin-care match of Candidate (4) and the more general match of Candidate (10). From a human perspective, the resulting trace is concise, coherent, and easy to follow, while avoiding much of the redundancy present in the raw CoT.

However, training the student on these summaries leads to substantially more parse failures. Most failures involve incomplete output lists, while duplicate indices are also considerably more common under LLM compression. We hypothesize that the highly structured, verbose summaries are out of distribution for the student and encourage it to imitate intermediate rankings, numbered explanations, product names, or other natural-language structures rather than reliably producing a single valid final list. Thus, although LLM summarization produces traces that appear high quality to human readers, they are less suitable as supervision for a smaller student model with strict output-format requirements.

\begin{promptbox}{LLM Summarized CoT}{olive}
<think>
Purchase History Analysis:
*   Purchase 1: Motions Naturally You, Hydrate My Curls Pudding (Hair care, curly hair, moisturizing/styling)
*   Purchase 2: Envyderm Eyelash Enhancement Serum (Eye/lash care, nighttime serum/growth)
*   Purchase 3: Hydroxatone Intensive Anti-Wrinkle Complex (Skin care, anti-aging)
*   Purchase 4: Tresemme Keratin Smooth Infusing Shampoo (Hair care, keratin/smoothing)
*   Purchase 5: Schick Hydro 5 Disposable Razor (Grooming, shaving)
*   Purchase 6: Schick Hydro Silk Disposable Razor (Grooming, shaving)

\medskip
User Preference Modeling:

* \textbf{Hair Care:} High interest in moisturizing, smoothing, and keratin-based products (P1, P4).
* \textbf{Skin Care:} Interest in anti-aging and specialized skin treatments (P3).
* \textbf{Eye/Lash Care:} Interest in lash enhancement (P2).
* \textbf{Grooming:} Interest in shaving/grooming tools (P5, P6).

\medskip
Candidate Analysis:

* Title: Organic Root Stimulator Olive Oil Replenishing Conditioner

  * Category: Hair Care (Conditioner, moisturizing/restorative)
  * Match level: High (Aligns with moisture themes in P1 and P4).
* Title: China Glaze Crackle Collection Lightening Bolt

  * Category: Nail Care (Nail polish)
  * Match level: Low (No nail care in history).
* Title: Tweezerman LTD Deluxe Metal Eyelash Curler

  * Category: Eye/Lash Care (Tool)
  * Match level: Medium-High (Direct category connection to P2).
* Title: L'Oreal Paris Age Perfect Hydra-Nutrition Moisturizer

  * Category: Skin Care (Anti-aging, moisturizing)
  * Match level: High (Direct theme match to anti-aging P3).
* Title: Burt's Bees Hair Repair Shea \& Grapefruit Deep Conditioner

  * Category: Hair Care (Conditioner, moisturizing)
  * Match level: High (Aligns with moisture themes in P1 and P4).
* Title: Cristophe Professional Glossing Shampoo

  * Category: Hair Care (Shampoo, smoothing/shine)
  * Match level: High (Direct type match to P4).
* Title: Essie Nail Lacquer, Midnight Cami

  * Category: Nail Care (Nail polish)
  * Match level: Low (No nail care in history).
* Title: Remington S8500 Shine Therapy Hair Straightener

  * Category: Hair Care (Tool, smoothing)
  * Match level: High (Thematic connection to smoothing in P4).
* Title: Manic Panic Pillarbox Red

  * Category: Hair Care (Hair dye)
  * Match level: Medium (Hair care category, but a different sub-category than smoothing/hydrating).
* Title: Dermalogica PreCleanse

  * Category: Skin Care (Cleansing oil)
  * Match level: Medium-High (General skin care connection to P3).

\medskip
Initial Ranking:

1. Candidates 1, 4, 5, 6 (Directly matching primary categories: Hair Care and Skin Care).
2. Candidate 3 (Eye care - related to P2).
3. Candidate 8 (Hair styling tool - related to hair care).
4. Candidate 10 (Skin care - related to P3).
5. Candidate 9 (Hair care - dye, less relevant than smoothing/hydrating).
6. Candidates 2, 7 (Nail care - no history).

\medskip
Verification/ Self-Correction:
\textbf{Refining Hair Care Ranking:}
The user has significant history in hair care (P1, P4).

* Candidate 6 (Shampoo) is a direct product-type match to P4 (Shampoo).
* Candidates 1 and 5 (Conditioners) match the moisturizing/replenishing themes of P1 and P4.
* Candidate 9 (Dye) and Candidate 8 (Straightener) are hair-related but are different sub-categories (color vs. tool).
  *Refinement:* Prioritize C6, then C1 and C5.

\medskip
\textbf{Refining Skin Care Ranking:}

* Candidate 4 (Anti-aging moisturizer) is a very strong match for the specific anti-aging interest in P3.
* Candidate 10 (Cleansing oil) is a general skin care product.
  *Refinement:* C4 should rank higher than C10.

\medskip
\textbf{Refining Tools and Other Categories:}

* Candidate 3 (Eyelash curler) matches the eye care category of P2.
* Candidate 8 (Hair straightener) matches the smoothing theme of P4.
* Comparing C3 vs C8: C3 is a direct category match (Eye care), whereas C8 is a thematic tool match. I will place C3 higher as it matches a specific purchase category.
* Candidates 2 and 7 (Nail care) have no historical relevance and remain at the bottom.

\medskip
\textbf{Final Ranking Order Construction:}

1. \textbf{Candidate 6} (Shampoo - direct type match to P4)
2. \textbf{Candidate 1} (Conditioner - moisture match to P1/P4)
3. \textbf{Candidate 5} (Conditioner - moisture match to P1/P4)
4. \textbf{Candidate 4} (Anti-aging moisturizer - direct theme match to P3)
5. \textbf{Candidate 3} (Eyelash tool - direct category match to P2)
6. \textbf{Candidate 8} (Hair tool - thematic match to P4)
7. \textbf{Candidate 10} (Skin cleanser - skin care category match)
8. \textbf{Candidate 9} (Hair dye - hair care category match)
9. \textbf{Candidate 2} (Nail polish - no match)
10. \textbf{Candidate 7} (Nail polish - no match)

</think>

[6, 1, 5, 4, 3, 8, 10, 9, 2, 7]
\end{promptbox}


\paragraph{SCOReD-Optimized CoT:} The SCOReD-optimized CoT quickly arrives at the final ranking, (\texttt{[6,1,5,4,3,8,10,9,2,7]}), and retains only the key comparisons needed to justify this ordering. It explains why Candidate (6) is ranked first, why Candidate (4) is the strongest skin-care match, and why Candidate (3) is placed above Candidate (8). This removes much of the redundant rechecking while preserving the reasoning that determines the final answer.

\begin{promptbox}{Our Compressed CoT}{ForestGreen}
<think>
\textcolor{gray}{$\triangleright$ Purchase History Analysis}
*   Purchase 1: Motions Naturally You, Hydrate My Curls Pudding (Hair care, curly hair, moisturizing)
    *   Purchase 2: Envyderm Eyelash Enhancement Serum (Eye/lash care, nighttime serum)
    *   Purchase 3: Hydroxatone Intensive Anti-Wrinkle Complex (Anti-aging skin care)
    *   Purchase 4: Tresemme Keratin Smooth Infusing Shampoo (Hair care, keratin/smoothing)
    *   Purchase 5: Schick Hydro 5 Disposable Razor (Shaving/Grooming)
    *   Purchase 6: Schick Hydro Silk Disposable Razor (Shaving/Grooming)

\medskip
\textcolor{gray}{$\triangleright$ User Preference / Theme Modeling}
*Key themes: Hair Care (moisturizing, smoothing, keratin), Skin Care (anti-aging, moisturizing), Eye Care (lashes), Grooming (shaving).*

\medskip
\textcolor{gray}{$\triangleright$ Candidate Analysis}
*   Title: Organic Root Stimulator Olive Oil Replenishing Conditioner
        *   Category: Hair Care (Conditioner, moisturizing/restorative)
        *   Match level: High (Hair care, similar to P1, P4).

*   Title: China Glaze Crackle Collection Lightening Bolt
    *   Category: Nail Care (Nail polish)
    *   Match level: Low (No nail care history).

*   Title: Tweezerman LTD Deluxe Metal Eyelash Curler
    *   Category: Eye/Lash Care (Tool)
    *   Match level: Medium-High (Eye care connection to P2).

*   Title: L'Oreal Paris Age Perfect Hydra-Nutrition Moisturizer
    *   Category: Skin Care (Anti-aging, moisturizing)
    *   Match level: High (Skin care, similar to P3).
*   Title: Burt's Bees Hair Repair Shea \& Grapefruit Deep Conditioner
    *   Category: Hair Care (Conditioner, moisturizing)
    *   Match level: High (Hair care, similar to P1, P4).
*   Title: Cristophe Professional Glossing Shampoo
    *   Category: Hair Care (Shampoo, smoothing/shine)
    *   Match level: High (Hair care, similar to P4).

*   Title: Essie Nail Lacquer, Midnight Cami
        *   Category: Nail Care (Nail polish)
        *   Match level: Low (No nail care in history).

*   Title: Remington S8500 Shine Therapy Hair Straightener; Category: Hair Care (Smoothing tool); Match level: High (Hair care connection to P4, P1).
*   Title: Manic Panic Pillarbox Red; Category: Hair Care (Hair dye); Match level: Medium (Hair care, but dye is different from smoothing/hydrating).

*   Title: Dermalogica PreCleanse
        *   Category: Skin Care (Cleansing oil)
        *   Match level: Medium-High (Skin care, similar to P3).
        
\medskip
\textcolor{gray}{$\triangleright$ Verification / Pairwise Comparison / Self-Correction}
* Analyze the thematic connections between preferences and candidates to refine the ranking:
    * P1 (Hydrate Curls Pudding) and P4 (Keratin Smooth Shampoo) focus on moisture and smoothing, matching Candidate 6 (Glossing Shampoo), Candidate 1 (Olive Oil Replenishing Conditioner), Candidate 5 (Shea \& Grapefruit Deep Conditioner), and Candidate 8 (Hair straightener).
    * P3 (Anti-wrinkle) matches Candidate 4 (Age Perfect Hydra-Nutrition Moisturizer) and Candidate 10 (Cleansing oil).
    * P2 (Eyelash Serum) matches Candidate 3 (Eyelash curler).
    * Candidate 9 (Hair dye) relates to hair care, while Candidates 2 and 7 (Nail polish) have no specific match.
    * Based on these thematic groupings (Hair Care > Skin Care > Lash Care > Hair Tool > Hair Dye > Nail Care), the refined ranking order is:
    1. Candidate 6 (Shampoo, direct type match to P4)
    2. Candidate 1 (Conditioner, moisture match to P1, P4)
    3. Candidate 5 (Conditioner, moisture match to P1)
\medskip
\textcolor{gray}{$\triangleright$ Verification / Pairwise Comparison / Self-Correction}
Should Candidate 8 (hair straightener) be higher? It is a tool, whereas the user bought hair care products. Candidate 3 (eyelash curler) is a tool for an eye product they purchased. Note that Candidate 4 (Skin) is extremely similar to P3 (Skin).

\medskip
\textcolor{gray}{$\triangleright$ Ranking Construction / Tiering}

*Finalizing the order:*
    [6, 1, 5, 4, 3, 8, 10, 9, 2, 7]
\medskip
\textcolor{gray}{$\triangleright$ Verification / Pairwise Comparison / Self-Correction}
Categorize and rank products by type to identify the best candidates:
* Hair care (P1, P4, P5, P6): The top priorities are Candidate 6 (Cristophe Glossing Shampoo), which is a very close match for P4 (Tresemme Keratin Smooth Shampoo), followed by Candidate 1 (Root Stimulator Conditioner) and Candidate 5 (Burt's Bees Shea \& Grapefruit Conditioner), where Candidate 5 may be a better moisturizing match for P1 (Motions Hydrate My Curls Pudding) than Candidate 1. Other hair candidates include 9 (Hair Dye) and 8 (Hair Tool/Straightener); while Candidate 8 shares the "smoothing" theme of P4, it is a lower priority than consumables. The hair ranking is 6, 1, 5, 9, 8.
* Skin care (P3): Candidate 4 (Age Perfect Hydra-Nutrition Moisturizer) is a strong match for P3 (Hydroxatone Intensive Anti-Wrinkle Complex), making it a higher priority than Candidate 10 (Dermalogica PreCleanse).
* Lash care (P2): Candidate 3 is an eyelash curler
\medskip
\textcolor{gray}{$\triangleright$ Ranking Construction / Tiering}
Final list: 6 (Shampoo), 1 (Conditioner), 5 (Conditioner), 4 (Skin moisturizer), 3 (Lash tool), 8 (Hair tool), 10 (Skin cleanser), 9 (Hair dye), 2 (Nail polish), 7 (Nail polish).
\medskip
\textcolor{gray}{$\triangleright$ Verification / Pairwise Comparison / Self-Correction}
Candidates 1 and 5 are both conditioners, while 6 is the only shampoo, making 6 the closest match to P4. For the remaining order, Candidate 4 (moisturizer) is a better use-case match for P3 (anti-wrinkle complex) than Candidate 3 (eyelash curler) is for P2 (eyelash serum). 
\medskip
\textcolor{gray}{$\triangleright$ Final Ranking Construction}
Final sequence: [6, 1, 5, 4, 3, 8, 10, 9, 2, 7].
</think>

[6, 1, 5, 4, 3, 8, 10, 9, 2, 7]    
\end{promptbox}

\section{Correctly Parsed Subset Results}
\label{app:correct-format-comp}

The all-sample evaluation penalizes malformed rankings by assigning them a score of zero. To separate ranking quality from output-format reliability, Table~\ref{tab:rerank-large-test-common} evaluates the $3{,}834$ test samples for which all five methods produce valid rankings.

On this common-parsed subset, the three trained students perform similarly. Their NDCG scores range from $0.8019$ to $0.8035$, and the difference between methods is within the margin of error. SCOReD obtains the highest NDCG, MRR, and MAP, while LLM summarization performs best on several Hit, Precision, and Recall metrics.

These small differences contrast with the much larger gap in the all-sample evaluation. In particular, the all-sample NDCG difference between SCOReD and LLM summarization is $0.0566$. This indicates that the primary weakness of one-shot LLM compression is its substantially higher probability of producing an unusable final output.

\begin{table}[!h]
\centering
\caption{Reranking results on the $3{,}834$ test samples, corresponding to $85.7\%$ of the test set, for which all five methods produce valid rankings. \textbf{Bold} and \underline{underlining} denote the best and second-best results among trained students, respectively.}
\label{tab:rerank-large-test-common}
\begin{tabular}{lrrr|rr}
\toprule
 & \multicolumn{3}{c}{\textbf{Trained SLMs}}
 & \multicolumn{2}{c}{\textbf{Reference LLMs}} \\
\cmidrule(lr){2-4}
\cmidrule(lr){5-6}
 & \textbf{Baseline}
 & \textbf{LLM-Summ.}
 & \textbf{SCOReD}
 & \textbf{Gemma-4-26B}
 & \textbf{Qwen-3.6} \\
\midrule
Model size
    & 0.6B & 0.6B & 0.6B & 26B-E4B & 35B-A3B \\
Training
    & SFT & SFT & SFT & --- & --- \\
\midrule
Common samples
    & \multicolumn{5}{c}{3834} \\
Trace length (K chars)
    & 8.4$\pm$1.9
    & 4.2$\pm$0.7
    & 6.2$\pm$1.7
    & 9.9$\pm$3.7
    & 9.4$\pm$2.0 \\
\midrule
NDCG
    & 0.8019
    & \underline{0.8031}
    & \textbf{0.8035}
    & 0.8047
    & 0.8049 \\
MRR
    & 0.7823
    & \underline{0.7851}
    & \textbf{0.7858}
    & 0.7851
    & 0.7858 \\
MAP
    & 0.6756
    & \underline{0.6769}
    & \textbf{0.6773}
    & 0.6807
    & 0.6806 \\
Hit@1
    & 0.6523
    & \textbf{0.6575}
    & \underline{0.6557}
    & 0.6599
    & 0.6599 \\
Hit@3
    & 0.8954
    & \underline{0.8978}
    & \textbf{0.9035}
    & 0.8944
    & 0.8925 \\
Hit@5
    & 0.9731
    & \textbf{0.9773}
    & \underline{0.9744}
    & 0.9690
    & 0.9684 \\
Recall@1
    & 0.2174
    & \textbf{0.2192}
    & \underline{0.2186}
    & 0.2200
    & 0.2200 \\
Recall@3
    & 0.5500
    & \underline{0.5519}
    & \textbf{0.5531}
    & 0.5555
    & 0.5555 \\
Recall@5
    & 0.7334
    & \textbf{0.7371}
    & \underline{0.7368}
    & 0.7391
    & 0.7372 \\
Precision@1
    & 0.6523
    & \textbf{0.6575}
    & \underline{0.6557}
    & 0.6599
    & 0.6599 \\
Precision@3
    & 0.5500
    & \underline{0.5519}
    & \textbf{0.5531}
    & 0.5555
    & 0.5555 \\
Precision@5
    & 0.4400
    & \textbf{0.4423}
    & \underline{0.4421}
    & 0.4435
    & 0.4423 \\
\bottomrule
\end{tabular}%

\end{table}

\clearpage

\section{Additional Training Dynamics}
\label{app:training-dynamics}

\begin{figure}[!h]
    \centering
    \begin{subfigure}[b]{0.48\textwidth}
        \centering
        \includegraphics[width=\textwidth]
        {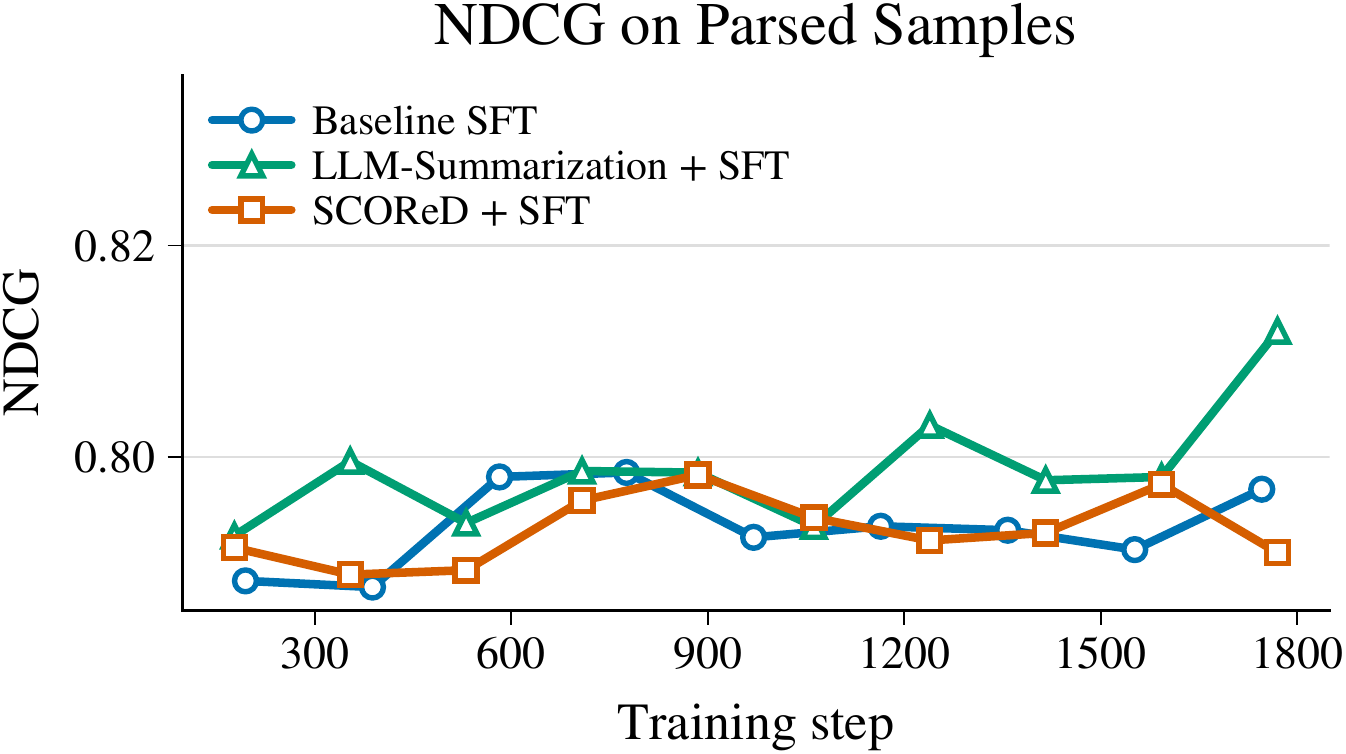}
        \caption{NDCG on correctly parsed samples.}
        \label{fig:ndcg-parsed-training}
    \end{subfigure}
    \hfill
    \begin{subfigure}[b]{0.48\textwidth}
        \centering
        \includegraphics[width=\textwidth]
        {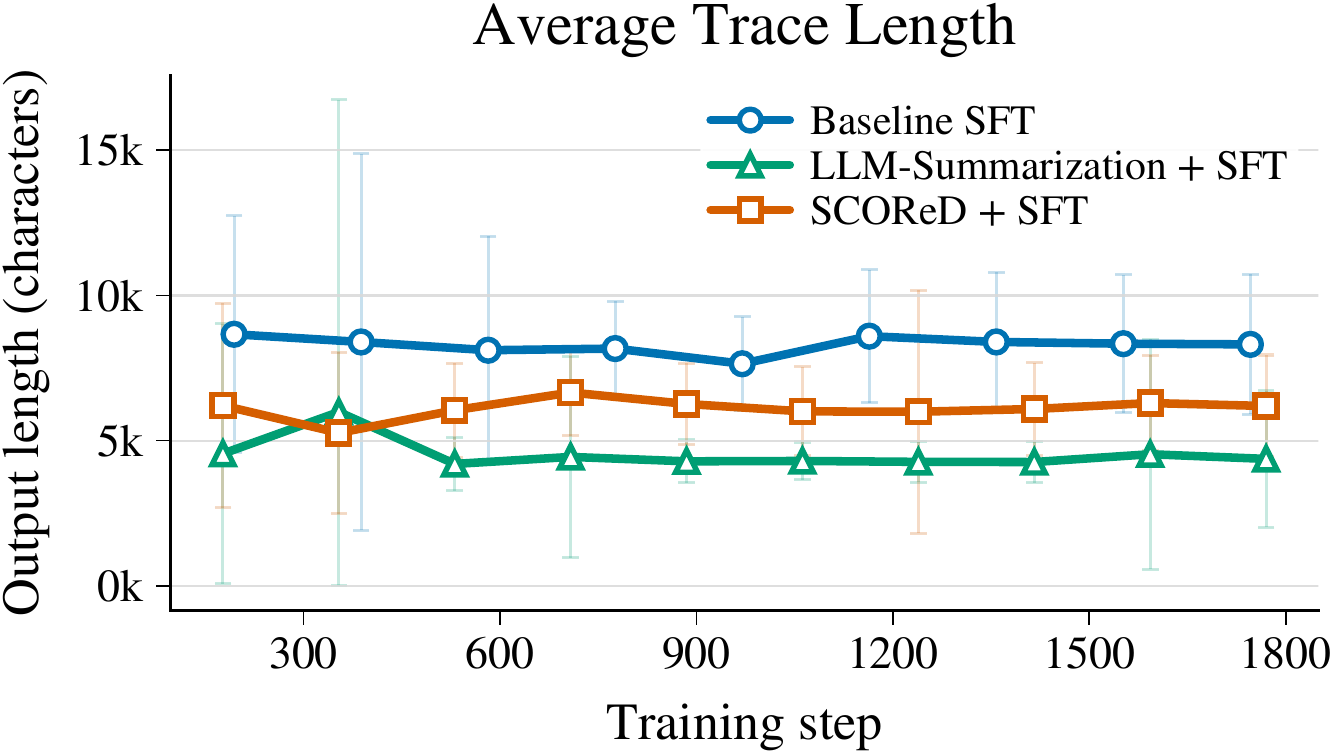}
        \caption{Average trace length.}
        \label{fig:output-length-training}
    \end{subfigure}
    \caption{Additional training dynamics for the three trained students.}
    \label{fig:appendix-training-dynamics}
\end{figure}

Figure~\ref{fig:appendix-training-dynamics} reports NDCG on correctly parsed samples and average trace length over training. Once malformed outputs are excluded, the ranking performance of the three students is substantially closer than in the all-sample evaluation. This further indicates that output-format reliability accounts for a large portion of the difference between the compression strategies.

The trace-length trajectories show that one-shot LLM summarization consistently produces the shortest outputs, while standard SFT produces the longest. SCOReD remains between these two extremes, reducing trace length without incurring the large parse-failure rate of the one-shot summarizer.

\begin{figure}[!h]
    \centering
    \includegraphics[width=0.5\linewidth]
    {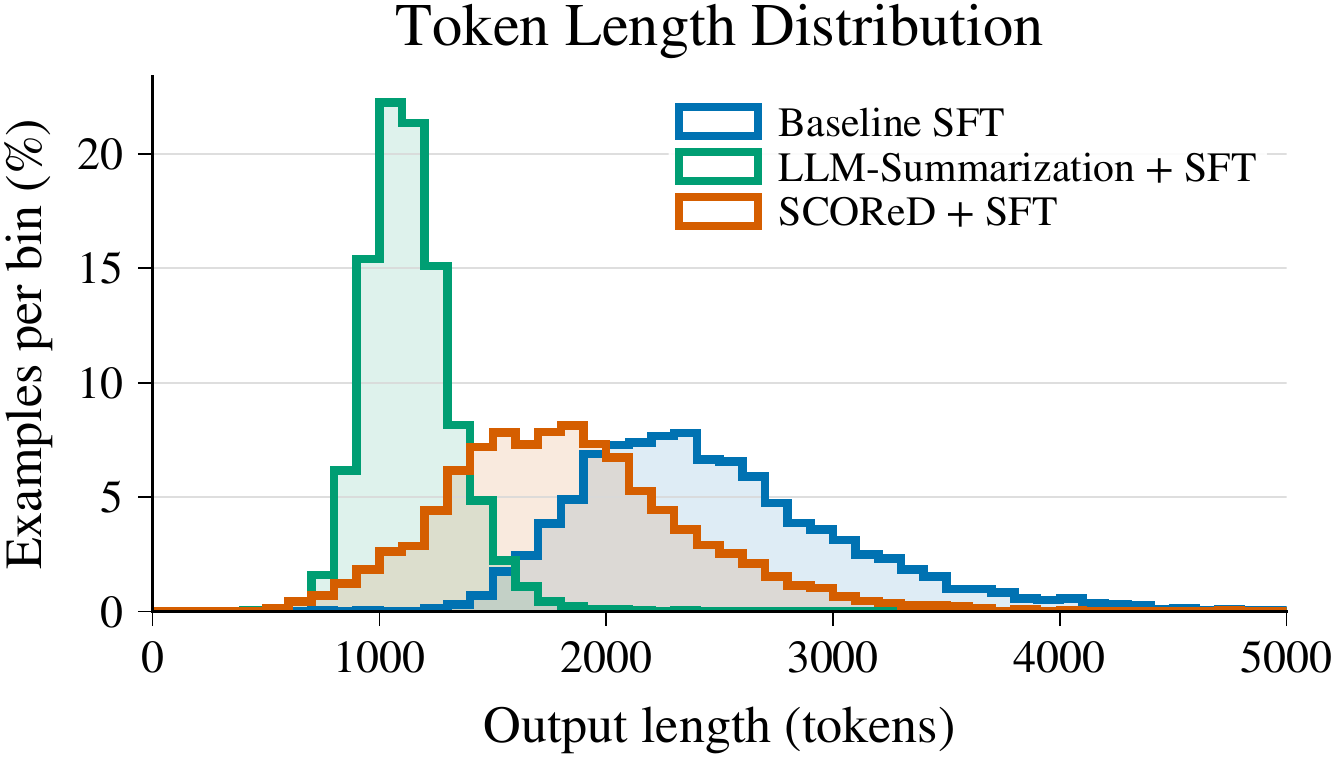}
    \caption{Distribution of student reasoning-trace lengths on the reranking test set.}
    \label{fig:trace-length-distribution}
\end{figure}

Figure~\ref{fig:trace-length-distribution} presents the test-time distribution of output lengths. Standard SFT has the longest and widest distribution. One-shot LLM summarization shifts the distribution toward substantially shorter traces, whereas SCOReD produces an intermediate and more concentrated distribution. Together with the all-sample results, this shows that the strongest compression ratio does not necessarily yield the most effective distillation traces.

\clearpage

\section{DAPO Results}
\label{app:dapo-results}

\begin{figure}[!h]
    \centering
    \includegraphics[width=0.4\linewidth]
    {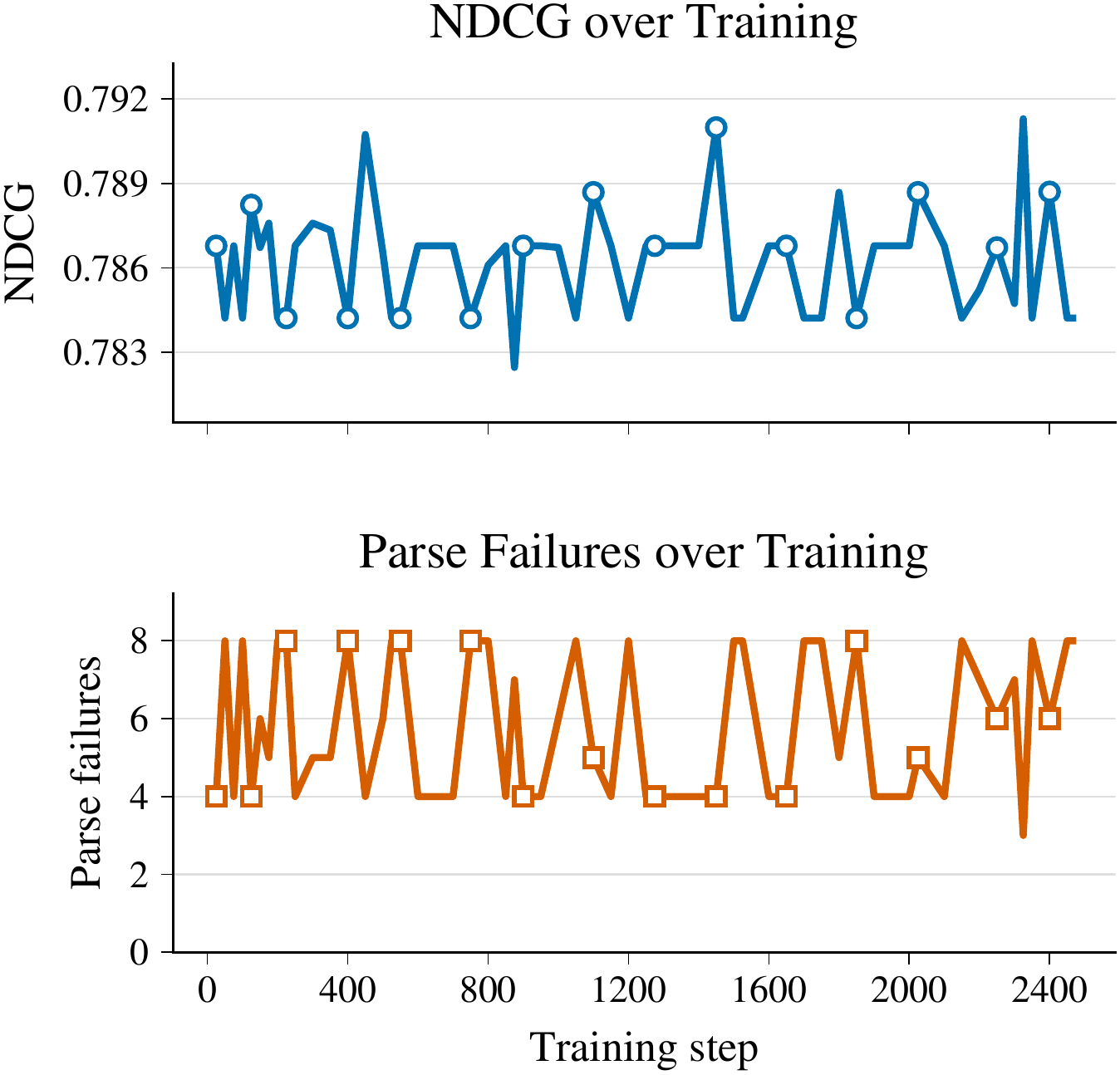}
    \caption{DAPO training over six epochs using NDCG maximization as the reward.}
    \label{fig:dapo-training}
\end{figure}

Figure~\ref{fig:dapo-training} shows DAPO training for six epochs using NDCG as the reward. NDCG oscillates within a narrow range and does not exhibit a sustained upward trend. Parse failures similarly fluctuate throughout optimization without a consistent reduction.

These results indicate that direct reinforcement learning does not improve upon the SCOReD SFT initialization. The limited movement in NDCG is consistent with the small remaining performance gap between the student and teacher: most sampled rankings are already similar in quality, reducing the frequency and magnitude of informative advantages.

\section{OPSD Results}
\label{app:opsd-results}

\begin{table}[!h]
\centering
\small
\caption{OPSD results on all test samples using privileged information. Checkpoint $0$ is the SCOReD SFT initialization.}
\label{tab:opsd-rec-full-test}
\begin{tabular}{lccccc}
\toprule
\textbf{Metric / Checkpoint}
    & \textbf{0 (SFT)}
    & \textbf{25}
    & \textbf{50}
    & \textbf{75}
    & \textbf{100} \\
\midrule
Training samples
    & 0 & 800 & 1600 & 2400 & 3200 \\
\midrule
NDCG
    & \textbf{0.7923}
    & 0.7908
    & 0.7898
    & 0.7907
    & 0.7878 \\
MRR
    & \textbf{0.7722}
    & 0.7721
    & 0.7702
    & 0.7719
    & 0.7689 \\
MAP
    & \textbf{0.6691}
    & 0.6667
    & 0.6660
    & 0.6667
    & 0.6639 \\
Hit@1
    & 0.6444
    & \textbf{0.6451}
    & 0.6417
    & 0.6442
    & 0.6419 \\
Hit@3
    & 0.8847
    & 0.8831
    & \textbf{0.8858}
    & 0.8809
    & 0.8804 \\
Hit@5
    & 0.9555
    & \textbf{0.9569}
    & 0.9564
    & \textbf{0.9569}
    & 0.9535 \\
Recall@1
    & 0.2148
    & \textbf{0.2150}
    & 0.2139
    & 0.2147
    & 0.2140 \\
Recall@3
    & \textbf{0.5470}
    & 0.5419
    & 0.5432
    & 0.5450
    & 0.5417 \\
Recall@5
    & \textbf{0.7269}
    & 0.7252
    & 0.7245
    & 0.7240
    & 0.7202 \\
Parse failure rate (\%)
    & \textbf{1.3858}
    & 1.4528
    & 1.5646
    & 1.4975
    & 1.8105 \\
\bottomrule
\end{tabular}%

\end{table}

\begin{table}[!h]
\centering
\small
\caption{OPSD results on the fixed correctly parsed subset of $4{,}147$ test samples. Checkpoint $0$ is the SCOReD SFT initialization.}
\label{tab:opsd-rec-full-test-common-parsed}

\begin{tabular}{lccccc}
\toprule
\textbf{Metric / Checkpoint}
    & \textbf{0 (SFT)}
    & \textbf{25}
    & \textbf{50}
    & \textbf{75}
    & \textbf{100} \\
\midrule
Training samples
    & 0 & 800 & 1600 & 2400 & 3200 \\
\midrule
NDCG
    & \textbf{0.8036}
    & 0.8023
    & 0.8019
    & 0.8024
    & 0.8015 \\
MRR
    & 0.7835
    & 0.7831
    & 0.7824
    & \textbf{0.7836}
    & 0.7819 \\
MAP
    & \textbf{0.6787}
    & 0.6763
    & 0.6758
    & 0.6762
    & 0.6751 \\
Hit@1
    & 0.6537
    & 0.6525
    & 0.6511
    & \textbf{0.6540}
    & 0.6520 \\
Hit@3
    & 0.8985
    & 0.8978
    & \textbf{0.9009}
    & 0.8944
    & 0.8951 \\
Hit@5
    & 0.9694
    & \textbf{0.9723}
    & 0.9718
    & 0.9715
    & 0.9718 \\
Recall@1
    & 0.2179
    & 0.2175
    & 0.2170
    & \textbf{0.2180}
    & 0.2173 \\
Recall@3
    & \textbf{0.5561}
    & 0.5505
    & 0.5517
    & 0.5529
    & 0.5503 \\
Recall@5
    & \textbf{0.7371}
    & 0.7363
    & 0.7352
    & 0.7346
    & 0.7335 \\
\bottomrule
\end{tabular}%

\end{table}

Table~\ref{tab:opsd-rec-full-test} reports OPSD performance on the complete test set. The SFT initialization at checkpoint $0$ achieves the highest NDCG, MRR, MAP, Recall@$3$, and Recall@$5$, as well as the lowest parse-failure rate. Although individual later checkpoints achieve small improvements on isolated Hit@$k$ metrics, these improvements are not consistent across metrics or checkpoints. By checkpoint $100$, NDCG decreases from $0.7923$ to $0.7878$, while the parse-failure rate increases from $1.39\%$ to $1.81\%$.

Table~\ref{tab:opsd-rec-full-test-common-parsed} shows the same comparison on the fixed correctly parsed subset. The differences remain small and non-monotonic. The initial checkpoint retains the highest NDCG, MAP, Recall@$3$, and Recall@$5$, while later checkpoints produce isolated improvements in MRR and Hit@$k$. OPSD therefore does not provide a systematic improvement over the SCOReD SFT initialization.

\end{document}